\documentclass[twocolumn,showpacs,aps,prb,superscriptaddress,floatfix, table]{revtex4-1}% Physical Review B

 \usepackage{array} %sets up table
\newcolumntype{L}[1]{>{\raggedright\let\newline\\\arraybackslash\hspace{0pt}}m{#1}}
\newcolumntype{C}[1]{>{\centering\let\newline\\\arraybackslash\hspace{0pt}}m{#1}}
\newcolumntype{R}[1]{>{\raggedleft\let\newline\\\arraybackslash\hspace{0pt}}m{#1}}
 
\usepackage{multirow}

\usepackage{graphicx}% Include figure files
\usepackage{dcolumn}% Align table columns on decimal point
\usepackage{bm}% bold math
\usepackage{topcapt}
\usepackage{amsmath}
\usepackage{color, colortbl}
\usepackage[usenames,dvipsnames]{xcolor}
\usepackage{xcolor}

\definecolor{maroon}{cmyk}{0,0.87,0.68,0.32}

\usepackage{placeins}
\definecolor{Gray}{gray}{0.7}

\newcommand{\fepo}{Fe$_3$PO$_4$O$_3$}
\newcommand{\fexpo}{Fe$_{(3-x)}$Ga$_x$PO$_4$O$_3$}

\begin{document}

\preprint{}

\title{Nanosized helical magnetic domains in strongly frustrated Fe$_3$PO$_4$O$_3$}

\author{K.A. Ross}
\affiliation{Department of Chemistry, Colorado State University, Fort Collins, Colorado 80523-1872, U.S.A.}
\affiliation{Department of Physics, Colorado State University, Fort Collins, Colorado 80523-1872, U.S.A.}

\author{M.M. Bordelon} 
\affiliation{Department of Chemistry, Colorado State University, Fort Collins, Colorado 80523-1872, U.S.A.}

\author{G. Terho} 
\affiliation{Department of Chemistry, Colorado State University, Fort Collins, Colorado 80523-1872, U.S.A.}

%\author{C. Brown}
%\affiliation{NIST}
%
%\author{A. Huq}
%\affiliation{Chemistry and Engineering Materials Division, Oak Ridge National Laboratory, Oak Ridge, Tennessee, 37831-6475, U.S.A}

\author{J.R. Neilson} 
\affiliation{Department of Chemistry, Colorado State University, Fort Collins, Colorado 80523-1872, U.S.A.}

\bibliographystyle{prsty}

\begin{abstract}
\fepo \ forms a non-centrosymmetric lattice structure (space group $R3m$) comprising triangular motifs of Fe$^{3+}$ coupled by strong antiferromagnetic interactions ($|\Theta_{CW}| > 900$ K).  Neutron diffraction from polycrystalline samples shows that strong frustration eventually gives way to an ordered helical incommensurate structure below $T_N$ = 163 K, with the helical axis in the hexagonal $ab$ plane and a modulation length of $\sim$ 86 \AA.  The magnetic structure consists of an unusual needle-like correlation volume that extends past 900 \AA \ along the hexagonal $c$-axis but is limited to $\sim$ 70 \AA \ in the $ab$ plane, despite the three-dimensional nature of the magnetic sublattice topology.   The small in-plane correlation length, which persists to at least $T = T_N/40$, indicates a robust blocking of long-range order of the helical magnetic structure, and therefore stable domain walls, or other defect spin textures, must be abundant in \fepo.  Temperature dependent neutron powder diffraction reveals small negative thermal expansion below $T_N$.  No change in lattice symmetry is observed on cooling through $T_N$, as revealed by high resolution synchrotron X-ray diffraction. The previously reported reduced moment of the Fe$^{3+}$ ions ($S$=5/2), $\mu \sim 4.2 \ \mu_B$, is confirmed here through magnetization studies of a magnetically diluted solid solution series of compounds, \fexpo, and is consistent with the refined magnetic moment from neutron diffraction  4.14(2) $\mu_B$.  We attribute the reduced moment to a modified spin density distribution arising from ligand charge transfer in this insulating oxide.  \end{abstract}

\maketitle

%%%%%%%%%%%%
%     Introduction
%%%%%%%%%%%%

\section{\label{sec:level1}Introduction}

%
%\textcolor{blue}{a) Schematic illustration of the unit cell of \fepo \ in the hexagonal setting. b) Triangular unit of magnetic Fe$^{3+}$ (5-fold coordinated by O$^{2-}$), with a PO$_4$ group linked to one face. } \textcolor{blue}{change size of a-c labels, OR change to Jamie's figure} \textcolor{blue}{will have to update this caption} a) Sublattice structure of magnetic Fe$^{3+}$ in \fepo.  Colored (grayscale) outlines indicate the fractional coordinate ($z$) along the hexagonal $c$ axis (note that the overall phase along $c$ is unconstrained by the hexagonal setting of $R3m$.  Here we choose blue (light gray): $z = 0.075$, magenta (dark gray): $z = 0.409$, red (dark gray): $z= 0.743$).  b) The two types of near-neighbor magnetic exchange interactions are shown in red (dark gray, $J_1$) and blue (light gray, $J_{2}$) c) Premedial lattice points (located at the center of each triangle) form a simple-rhombohedral lattice. The rhombohedral setting of the unit cell for \fepo \ is shown for comparison as thick black lines.

\begin{figure}[!htb]  
\centering
\includegraphics[ width=\columnwidth]{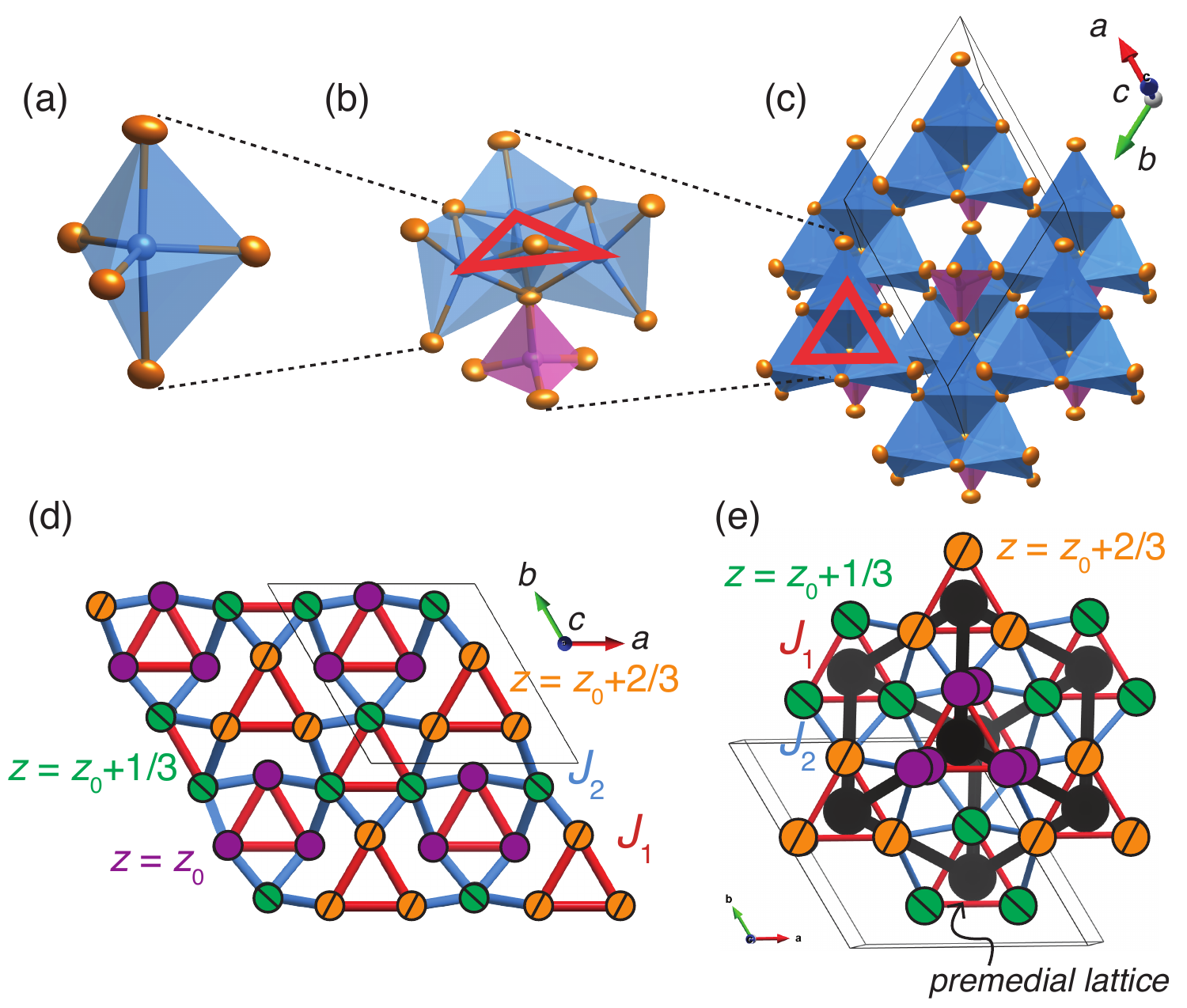}
\caption{Schematic illustrations of the crystal structure of \fepo, shown in the hexagonal setting of $R3m$.  (a) The local coordination of Fe is a distorted trigonal bipyramid, forming FeO$_5$ clusters.   (b) FeO$_5$ clusters are arranged in triangular subunits, with a PO$_4$ group linked to one face of the triangle.  (c)  The triangular units stack along $c$.    (d) Sublattice structure of magnetic Fe$^{3+}$ in \fepo, viewed along the $c$-axis.  Colored (grayscale) circles indicate the fractional coordinate ($z$) along the hexagonal $c$ axis.  Note that the overall phase along $c$ is unconstrained by $R3m$.  Here we choose purple (dark gray, no line): $z = 0.075$, green (medium gray, left diagonal line): $z = 0.409$, orange (light gray, right diagonal line): $z= 0.743$.  The two types of near-neighbor magnetic exchange interactions are shown in red (dark gray, $J_1$) and blue (medium gray, $J_{2}$) (e) Premedial lattice points (black spheres, located at the center of each triangular subunit) form a simple-rhombohedral lattice.}
\label{fig:structurefig}
\end{figure}

Materials that display helical magnetic order, particularly those that are non-centrosymmetric, are of current interest for their potential technological uses and fundamentally unique physical properties.\cite{nagaosa2013topological}  Materials with crystal structures lacking an inversion center, such as the B20-type metallic magnets MnSi,\cite{muhlbauer2009skyrmion} Fe$_{(1-x)}$Co$_x$Si,\cite{munzer2010skyrmion} and FeGe,\cite{yu2011near} as well as insulating Cu$_2$OSeO$_3$, \cite{seki2012observation} form helical (phase modulated) incommensurate magnetic ordered phases which can be converted, by applying a small magnetic field, into a regular lattice of vortex-like topological spin textures known as Skyrmions.  Once created, Skyrmions are stable against dissipation due to their topological nature.  Furthermore, they can be manipulated with electric currents \cite{jonietz2010spin} or thermal gradients, \cite{kong2013dynamics} making the transport of isolated Skyrmions promising as a new technology to realize high density and low dissipation spintronics devices.  Recent proposals for \emph{antiferromagnetic} (AFM) Skyrmions (those arising from an underlying AFM helical structure) have shown that these theoretically predicted objects may offer practical advantages by avoiding the Skyrmion hall effect. \cite{zhang2015antiferromagnetic, barker2015antiferromagnetic, zhang2015magnetic}   The role of defects in the ``parent'' helical structure appears to be essential in the generation of isolated Skyrmions, which can potentially be nucleated at sample edges, \cite{iwasaki2013current} at helical domain wall intersections,\cite{uchida2006real} and by combining helical domain walls. \cite{zhou2014reversible, zhang2015antiferromagnetic}

%\begin{figure}[!tb]  
%\centering
%\includegraphics[ width=\columnwidth]{./structure_fig_zero.pdf}
%\caption{a) Schematic illustration of the unit cell of \fepo \ in the hexagonal setting. b) Triangular unit of magnetic Fe$^{3+}$ (5-fold coordinated by O$^{2-}$), with a PO$_4$ group linked to one face. }
%\label{fig:structure_uc}
%\end{figure}

%\begin{figure*}[!tb]  
%\centering
%\includegraphics[width=2\columnwidth]{./structure_fig_one_v2.pdf}
%\caption{\textcolor{blue}{change size of a-c labels, OR change to Jamie's figure} \textcolor{blue}{will have to update this caption} a) Sublattice structure of magnetic Fe$^{3+}$ in \fepo.  Colored (grayscale) outlines indicate the fractional coordinate ($z$) along the hexagonal $c$ axis (note that the overall phase along $c$ is unconstrained by the hexagonal setting of $R3m$.  Here we choose blue (light gray): $z = 0.075$, magenta (dark gray): $z = 0.409$, red (dark gray): $z= 0.743$).  b) The two types of near-neighbor magnetic exchange interactions are shown in red (dark gray, $J_1$) and blue (light gray, $J_{2}$) c) Premedial lattice points (located at the center of each triangle) form a simple-rhombohedral lattice. The rhombohedral setting of the unit cell for \fepo \ is shown for comparison as thick black lines. } 
%\label{fig:structure_abstract}
%\end{figure*}

In order to understand magnetic defects of helical magnets and their manipulation, it is therefore of interest to explore systems in which helical spin order is susceptible to defect formation.   Helical order can be stabilized by the Dzyaloshinskii-Moryia (DM) interaction favoring non-coplanar spins, or alternatively through a balance of competing interactions, i.e., frustration. In frustrated systems, the ground state is usually selected from a plethora of closely competing phases; in principle this leads to flexibility in defect generation of helical structures.   For instance, substitution of non-magnetic atoms may allow for experimental control of defect formation, as interruptions in the connectivity of the magnetic sublattice can tip the delicate balance of frustrated interactions, potentially favoring commensurate structures or providing sites for defect nucleation.

In this report, we present a reinvestigation of the triangle-based material \fepo, known as the mineral Grattarolaite.\cite{cipriani1997rodolicoite}  We show that \fepo \ forms a high density of defects within an antiferromagnetic helical phase.  \fepo \ was first discovered and studied via magnetization,\cite{modaressi1983fe3po7,gavoille1987magnetic} specific heat,\cite{shi2013low} neutron powder diffraction,\cite{gavoille1987magnetic} and M\"ossbauer spectroscopy \cite{modaressi1983fe3po7, gavoille1987magnetic} several decades ago.  Its non-centrosymmetric lattice (space group $R3m$) consists of triangular units of Fe$^{3+}$ that are coplanar with the hexagonal $ab$ plane and are linked to a PO$_4$ group (Fig. \ref{fig:structurefig} a and b).   Along the $c$ axis, layers of triangular units arrange in a larger, triangular lattice pattern (Fig. \ref{fig:structurefig} c and d).   However, the structure is not quasi-two dimensional: interplane (along the hexagonal c axis) Fe-triangular units along the hexagonal $c$ axis are separated by 3.13 \AA, while the intraplane distance (in the $ab$ plane) is 4.78 \AA. In effect, the three-dimensional magnetic sublattice is best described by a decorated ``simple rhombohedral lattice," i.e., a simple cube compressed along the body diagonal, which forms the ``premedial lattice'' composed of sites located at the center of the triangular subunits (Fig. \ref{fig:structurefig} e).

Previous studies of \fepo \  reported a N\'eel transition at $T =164.5$ K via specific heat \cite{shi2013low} and  $T=173 \pm 5$ K via magnetic susceptibility. \cite{gavoille1987magnetic}  Above this transition, the paramagnetic susceptibility was fit using two forms of a Curie-Weiss law, each of which gave unusual results for the effective moment, as will be discussed below.  However, it is clear that the interactions are strong ($|\Theta_{CW}|$ $\sim$ 1000 K) and antiferromagnetic.   The frustration parameter, $|\Theta_{CW}| / T_N$, is therefore greater than 6 and indicates significant frustration of the antiferromagnetic interactions.  Below the transition, comparison of neutron powder diffraction at 4.2 K and at 200 K revealed the co-development of broad flat-topped peaks with a single sharp peak, all of which were assumed to be magnetic in origin. \cite{gavoille1987magnetic}  An AFM collinear commensurate magnetic structure was proposed, and was found to account for the \emph{central} position of each diffraction peak and its integrated intensity.  The broadening of select peaks was attributed to an anisotropic correlation volume, but this was not studied further.

In this contribution, we report new thermodynamic, synchrotron X-ray diffraction (SXRD), and neutron powder diffraction (NPD) measurements on \fepo. The high momentum ($Q$) resolution of our NPD patterns allows us to rule out the previously proposed commensurate magnetic structure. We show that a  \emph{helical incommensurate} structure with a strongly anisotropic correlation volume best describes this magnetic neutron diffraction pattern.  The model consists of magnetic domains restricted to 70 \AA \ in the hexagonal $ab$ plane but unrestricted along the $c$ axis; i.e., needle-like domains.  Based on the magnetic structure, we estimate the ratio of the two nearest neighbor exchange couplings, allowing for analysis of the energy cost for simple domain walls. Although the microscopic mechanism stabilizing the high density of domain walls remains unclear, we demonstrate that the domain walls are expected to form perpendicular to the $ab$ plane. 

The magnetic properties of the solid solution, \fexpo, in which non-magnetic Ga$^{3+}$ replaces Fe$^{3+}$, permits disambiguation of the size of the iron moment.   Curie-Weiss analysis of high temperature susceptibility of the \fexpo \ series provides a measure of the Fe$^{3+}$ moment, which is consistent with magnetic model obtained from NPD.  The moment is reduced by 16\% from the expected value for $S=5/2$, pointing to covalency (ligand charge transfer) effects.  Finally, the temperature dependence of the chemical structure is presented and shows some evidence for magneto-structural coupling, but without a symmetry-breaking structural phase transition occurring between 300 K and 4.5 K.

\section{\label{sec:level1}Experimental procedure}

Polycrystalline samples in the \fexpo \ series were synthesized by standard solid state methods, following previous Fe$_2$O$_3-$FePO$_4$ phase diagram reports. \cite{zhang2011phase} For the $x=0$ compound, appropriate stoichiometric amounts of dried Fe$_2$O$_3$ and FePO$_4$ were thoroughly mixed. Powders were pressed into pellets and sintered in alumina crucibles at 950\,$^\circ$C for 24 hours. Pellets were reground and heated multiple times at 1050\,$^\circ$C for 48 hours, until phase purity was maximized. Small amounts of Fe$_2$O$_3$ and FePO$_4$ (1$-$4 \% by mol) persisted even after three to four regrindings and reheatings.  If residual FePO$_4$ was detected (but only in the absence of Fe$_2$O$_3$), it was removed by overnight suspension in dilute HCl (2 M),  followed by rinsing with deionized water and drying in air.  

Phase identity and cell parameters were determined by refining powder X-ray diffraction patterns from a Scintag Advanced Diffraction System (Cu-K$\alpha$ radiation, $\lambda = $1.541 \AA) using the Rietveld method as implemented by GSAS/EXPGUI.\cite{toby2001expgui} The remaining members of the \fexpo \ series ($x = 0.5, 1.0, 1.5, 2.0, 2.5, 2.7$) were prepared and analyzed similarly with stoichiometric amounts of dry Fe$_2$O$_3$, Ga$_2$O$_3$, FePO$_4$, and GaPO$_4$. Samples of high Ga$-$content ($x = 2.0, 2.5, 2.7$) required more heatings to achieve maximal phase purity, and the \fexpo \ type structure was not the dominant phase at $x=2.9$ indicating a solubility limit between $x=2.8$ and 2.9 (See Appendix \ref{sec:fexpoxray}).  None of the solid-solution compounds were treated with HCl. 

The unit cell parameters, $a$ and $c$, for \fexpo \ were extracted from Rietveld refinements of powder X-ray diffraction patterns. These unit cell lengths decrease with increasing Ga(III) (ionic radius in 5-fold coordination$=55.0$\,pm)\cite{shannon1976revised} substitution for Fe(III) (ionic radius in 5-fold coordination $= 58.0$\,pm)\cite{shannon1976revised} (Fig. \ref{fig:fexpoxray}). As expected from Vegard's law, the decrease in lattice parameters is linear as a function of $x$.

Magnetization and specific heat measurements were performed using a Quantum Design Inc. Physical Properties Measurement System. The magnetization of \fepo \ and the gallium-diluted \fexpo \ series were measured from 1.8 K to 700\,K under zero-field cooled (ZFC) and field-cooled (FC) conditions at $\mu_0$H = 1 T. Specific heat measurements were carried out using the semi-adiabatic heat-pulse method at temperatures from 1.8 K to 300 K.

 Neutron powder diffraction (NPD) experiments were carried out at $T= [4.5,60, 100,120,160, 220, 295]$ K on a polycrystalline sample of \fepo \ ($m = $1.8 g) using the BT1 powder diffractometer at the NIST Center for Neutron Research.  A constant wavelength of 2.0799  \AA \ with 15' collimation was used, providing a resolution of $\delta Q$/$Q$ = 0.013 at $Q=1.0$ \AA.\cite{bt1}   A polycrystalline sample of \fepo \ ($m = $3.3 g) was studied at $T=$ 295 K and 100 K using the POWGEN time-of-flight neutron diffractometer at the Spallation Neutron Source, Oak Ridge National Laboratory.  The long-wavelength frame (central wavelength = 2.665 \AA) was used for magnetic structure refinements, providing a resolution of $\delta Q$/$Q$ = 0.005 at $Q=1.0$ \AA.\cite{POWGEN}   Synchrotron X-ray Diffraction (SXRD) data were collected on \fepo \ using the diffractometer on the 11-BM-B beam line at the Advanced Photon Source, Argonne National Laboratory at temperatures of $T$ = 100\,K and 295\,K using an incident wavelength of 0.414 \AA.\cite{wang2008dedicated} To account for X-ray absorption of the sample, a correction of $\mu$R = 0.40, calculated from the sample composition and radius, was applied to the full pattern analysis

\section{\label{sec:level1}Results}

\subsection{Thermodynamic measurements}

\begin{figure}[!tb]  
\centering
\includegraphics[ width=\columnwidth]{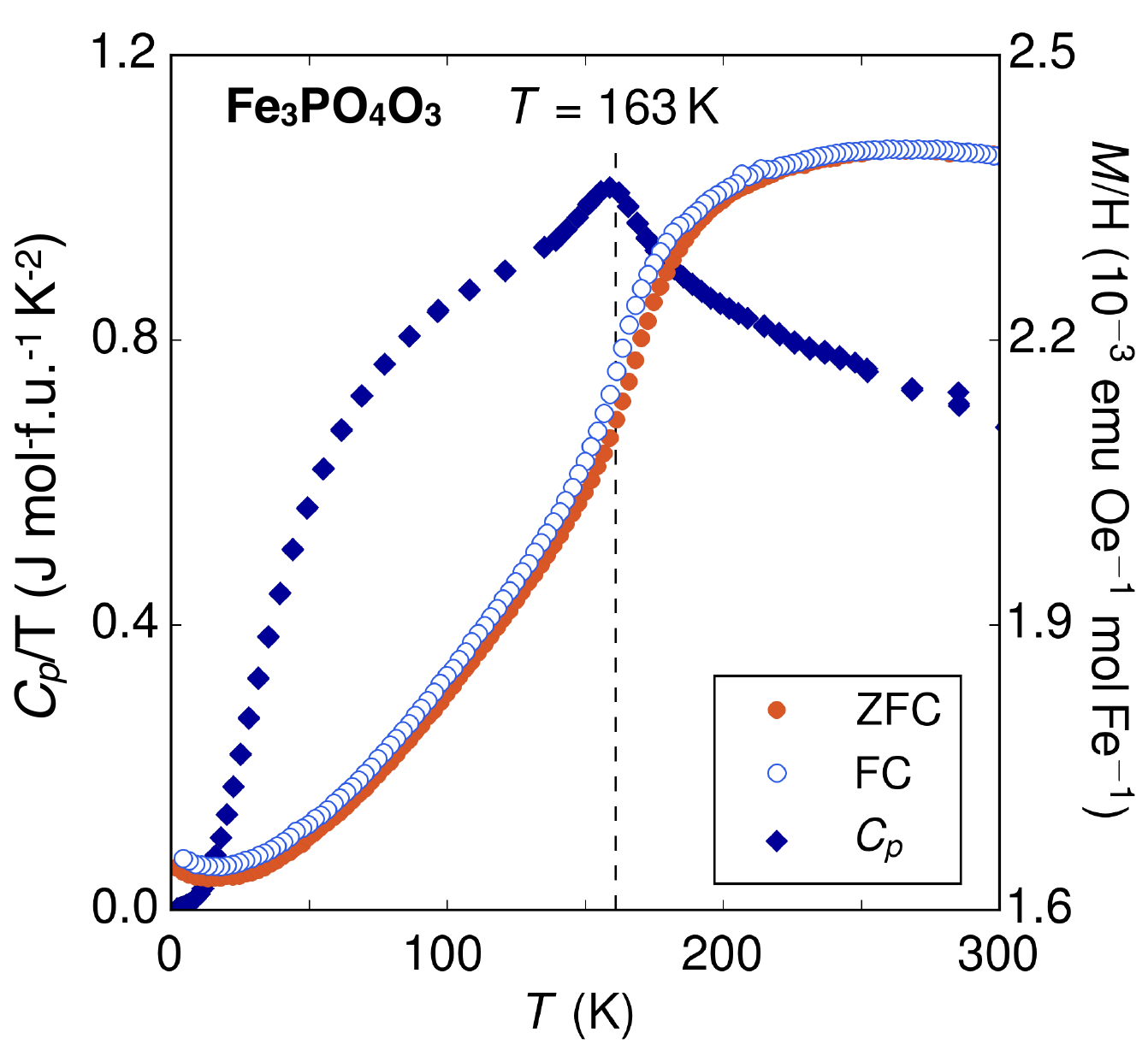}
\caption{ $C_p/T$ vs. $T$ (left axis) and $M/$H vs. $T$ (H $= 10000 $ Oe, right axis) for polycrystalline \fepo .  The data show a N\'eel transition at $T_N = 163$ K.  The field cooled (FC) vs. zero field cooled (ZFC) magnetization measurements are nearly identical over the full temperature range. }
\label{fig:Cp}
\end{figure}

\begin{figure}[!tb]  
\centering
\includegraphics[ width=0.8\columnwidth]{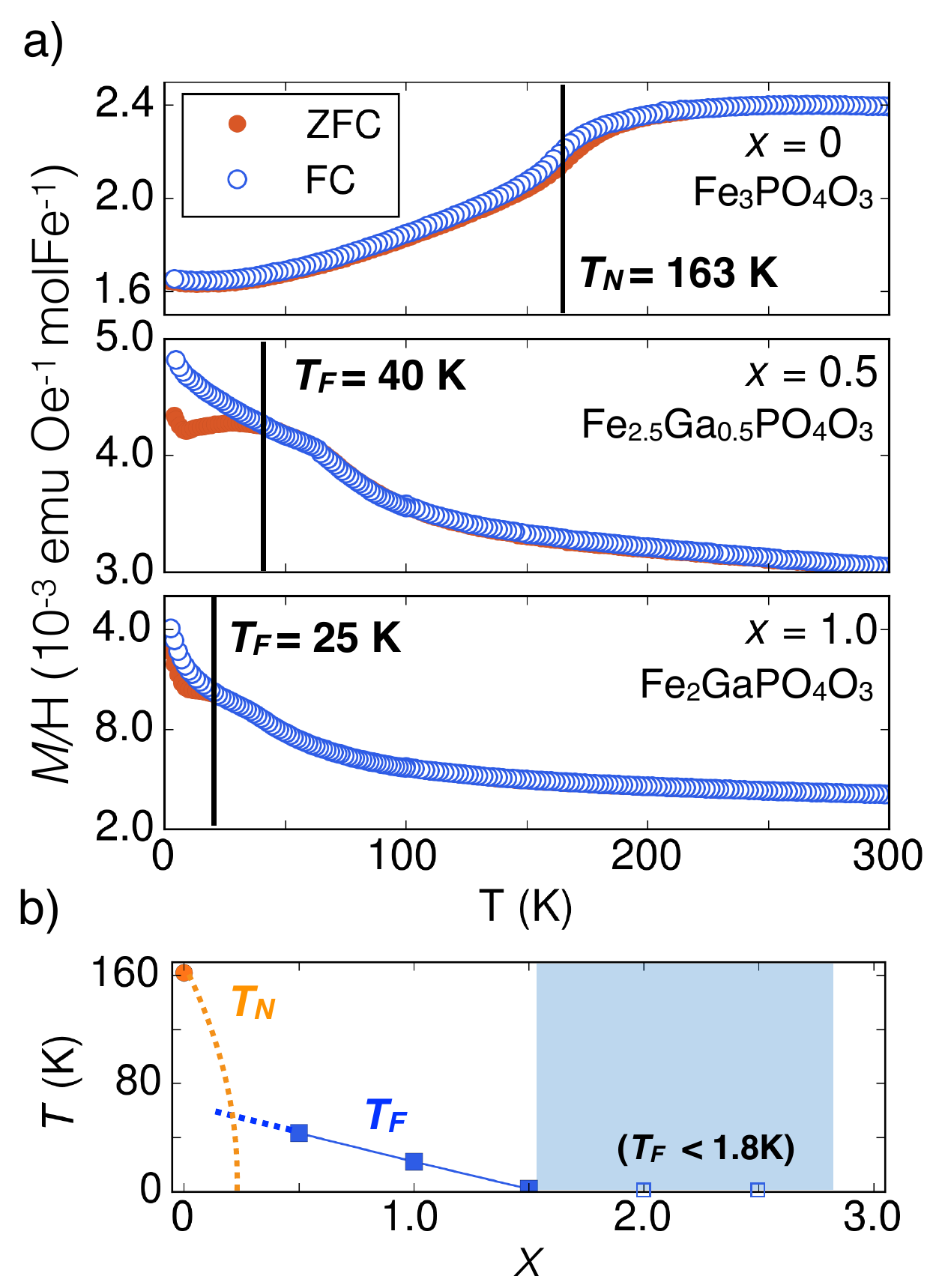}
\caption{ (a) Susceptibility measurements ($M$/H) made on representative members of the solid solution \fexpo \ with an applied field of 10000 Oe.  At $x=0$, i.e., \fepo, the FC and ZFC curves do not fully diverge, but show a small splitting around $T_N$.  At higher $x$, the FC vs. ZFC curves split at a temperature which decreases with increasing $x$, as might be expected for a spin glass.  The temperature of the splitting is referred to as $T_F$.  (b)   $T_N$ and $T_F$ as a function of gallium content ($x$) in \fexpo.  Solid line is a guide to the eye.  Dotted lines show possible trends of $T_N$ and $T_F$ as a function of $x$, but are not constrained by the data.  The blue (gray) region indicates composition range for which no distinguishing features are observed above $T=1.8$ K (based on data taken at the two compositions marked by open blue squares). }
\label{fig:lowTvsm}
\end{figure}

\subsubsection{Specific Heat}
\begin{figure}[!tb]  
\centering
\includegraphics[ width=\columnwidth]{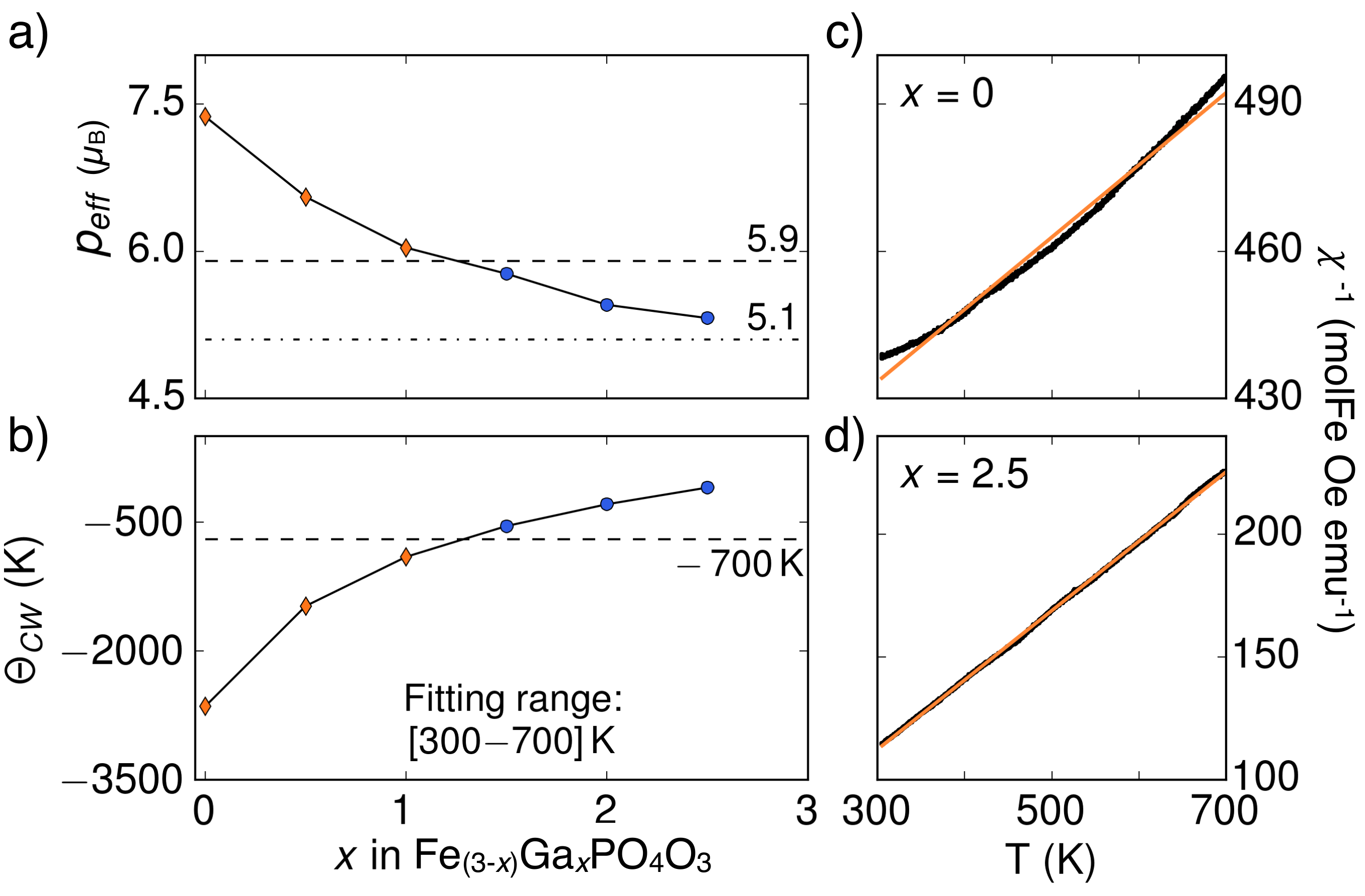}
\caption{ (a) and (b) Results of Curie-Weiss analysis of the inverse magnetic susceptibility of \fexpo.  Fitting was carried out in the temperature range of 300 K to 700 K.  For $x \ge 1.0$ (red diamonds), the analysis produces unreliable results due to non-linearity in $\chi^{-1}$.   At higher $x$, the magnetic dilution the inverse susceptibility data become linear over 300-700 K to allow reliable application of the Curie-Weiss relation.  As the range of compositions increases $x>1.0$, the effective moment tends towards $p_{\text{eff}}$  = 5.1 $\mu_B$, less than the value expected for $S=5/2$ (5.9).  (c) and (d) Inverse susceptibility at H $= 10000$ Oe for $x=0$ (c) and $x=2.5$ (d).  Orange lines are fits to the Curie-Weiss law. }
\label{fig:CW}
\end{figure}

The specific heat of \fepo \ was originally reported in Ref. \onlinecite{shi2013low}.  An anomaly was observed near 163 K, corresponding to the magnetic transition temperature.  We have confirmed this specific heat, shown in Fig. \ref{fig:Cp}.  In contrast to Shi \emph{et al}, we do not attempt to isolate the magnetic contribution to the specific heat.  Due to the presence of strong magnetic correlations above $T_N$, evidenced by both inverse susceptibility (Fig. \ref{fig:CW}) and NPD (Fig. \ref{fig:BT1}), estimating the lattice contribution is not accurate and appears to lead to spurious results such as the unphysically small change in entropy reported in Ref. \onlinecite{shi2013low} which is likely amplified over the broad temperature range of the anomaly in the specific heat.  Our attempts to synthesize the $x=3$ end member of \fexpo, Ga$_3$PO$_4$O$_3$, for use as a lattice analog were not successful.  

\subsubsection{Low Temperature Susceptibilty}
The anomaly observed in the specific heat of \fepo \ coincides with an inflection point in its sharply decreasing magnetization, confirming the formation of an ordered antiferromagnetic structure at 163 K (Fig. \ref{fig:Cp}).  The ZFC and FC susceptibility at $\mu_0$H $= 1$ T do not diverge appreciably, though there is a small gap between the two which develops above $T_N$, is maximal at $T_N$, and takes on a nearly constant value down to 1.8 K.   There is an upturn in the magnetization near 1.8 K which likely arises from ``orphan spins'' at lattice defects.

Members of the solid solution \fexpo \ show a dramatic change in the the behavior of their low temperature susceptibility compared to \fepo, as shown in Fig. \ref{fig:lowTvsm} for members $x = 0.5$ and $x = 1.0$.  At $x=0.5$, the AFM transition appears to be totally suppressed, but a transition which resembles freezing (due to the presence of a ZFC / FC splitting) occurs at $T_F$ = 40\,K.  This feature persists at higher $x$ but $T_F$ decreases; at Ga concentrations higher than $x=1.5$, any such features, if they exist, are outside of the temperature range of the measurement (i.e., $T_F<1.8$ K).  Fig. \ref{fig:lowTvsm} b) illustrates the magnetic phase evolution as a function of $x$ in the \fexpo \ system.

\subsubsection{High Temperature Susceptibility}
\label{sec:CW}
The magnetic susceptibility in \fepo \ and \fexpo \ measured from 300 K to 900 K were analyzed with the Curie-Weiss relation (the susceptibility for \fepo\ over the full measured temperature range, 1.8 K to 900 K, is shown in Appendix \ref{sec:fullchi}). Constants $C$ and $\Theta_{CW}$ were extracted (Table \ref{table:CW}) by fits to the inverse Curie-Weiss relation (Eqn. \ref{eqn:CW}) via the linear least-squares method from 300$-$700\,K. The effective paramagnetic moment, $p_{\text{eff}} = \sqrt{8\,C}$, was subsequently calculated using the extracted $C$ value.

\begin{equation}
	\chi^{-1} = \frac{(T - \Theta_{CW})}{C}
\label{eqn:CW}
\end{equation}

%!{\vrule width -1pt}
\begin{table}[h]
\caption{Results from magnetic susceptibility fits to the inverse Curie-Weiss equation (\ref{eqn:CW}) for \fepo \ and \fexpo \ from 300 $-$ 700\,K.  The effective moment is calculated as $p_{\text{eff}} = \sqrt{8\,C}$, where $C$ is the inverse slope of the fit in CGS units of emu$\cdot$K$^{-1}\cdot$molFe$^{-1}\cdot$Oe$^{-1}$. Negative values of $\Theta_{CW}$ indicate an overall antiferromagnetic interaction. }
\begin{tabular}{ C{1.6cm}  C{3cm}  C{1.6cm}  C{1.6cm} }
\hline \hline
$x$   & \begin{tabular}[c]{@{}c@{}}$C$ \\ \end{tabular} & \begin{tabular}[c]{@{}c@{}}p$_{eff}$ \\ ($\mu_B$)\end{tabular} & \begin{tabular}[c]{@{}c@{}}$\Theta_{CW}$ \\ (K)\end{tabular} \\ \hline
1.5   & 4.16                                                                   & 5.77                                                           & -545                                                         \\
2.0   & 3.72                                                                   & 5.45                                                           & -292                                                         \\
2.5   & 3.54                                                                   & 5.32                                                           & -97                                                          \\ \hline \hline
\end{tabular}
\label{table:CW}
\end{table}

 We find that samples of \fexpo \ from $x = 0.0$ to $x = 1.0$ display nonlinear inverse susceptibilities (Fig. \ref{fig:CW}) and are unfit for Curie-Weiss analysis.  Dilution with nonmagnetic Ga above 50\% substitution from $x = 1.5$ to $x = 2.7$ produces a paramagnetic response that is well described by the Curie-Weiss relation. The magnetic behavior of Fe atoms can be described by the extracted constants ($\Theta_{CW}$, $C$) and the relations $C = (p_{\text{eff}}/8)^2$, where $p_{\text{eff}} =  2\sqrt{S(S+1)}$ (Table \ref{table:CW}). As $x$ increases, the magnitude of $\Theta_{CW}$ decreases, reflecting the decrease in the antiferromagnetic Fe-Fe interactions. The Fe atom effective moment trends to $\sim$ 5.1 $\mu_B$ as $\Theta_{CW}$ falls below the lowest temperature used for fitting (Fig. \ref{fig:CW}).  Typical spin-only Fe$^{3+}$ ($S=5/2$) produces $p_{\text{eff}}  = 5.9 \mu_B$. For the members of the solid solution that are reliably described by the Curie-Weiss relation ($x \geq 1.5$), $p_{\text{eff}}$ is noticeably lower than the spin-only effective moment.

\begin{table*}[!ht]
\caption{Crystal structure parameters of \fepo \ (space group $R3m$) obtained from Rietveld analysis of 11$-$BM data (SXRD) at 100\,K and 295\,K. The $z-$position of the Fe-atom is held constant when refining atom positions (see text).  $U_{iso}$ is shown in units of 10$^{-3}$ \AA$^2$.}
\begin{tabular}{C{1.5cm} C{1.2cm} C{1.7cm} C{1.7cm} C{1.7cm} C{1.65cm} C{0.05cm} | C{1.7cm} C{1.7cm} C{1.7cm} C{1.7cm} }
\hline \hline
                           &     & \multicolumn{4}{c}{100\,K}                       &  & \multicolumn{4}{c}{295\,K}                       \\ \hline
\multirow{2}{*}{\begin{tabular}[c]{@{}c@{}}Unit Cell\\ Parameter\end{tabular}} & $a$ & \multicolumn{4}{c}{7.999(1) \AA}                 &  & \multicolumn{4}{c}{8.004(1) \AA}                 \\
                           & $c$ & \multicolumn{4}{c}{6.852(1) \AA}                 &  & \multicolumn{4}{c}{6.860(1) \AA}                 \\ \hline
Atom  & Wyckoff Position & $x$ & $y$ & $z$ & $U_{iso}$ &  & $x$ & $y$ & $z$ & $U_{iso}$ \\ \hline
Fe & 9b & 0.7967(1) & -0.7967(1) & 0.7425 & 1.63(2) &  & 0.7970(1) & -0.7970(1) & 0.7425 & 4.83(3) \\
P & 3a & 0 & 0 & 0.0015(1) & 1.79(12) &  & 0 & 0 & 0.0002(1) & 5.21(13) \\ %0.00017(9)
O1 & 3a & 0 & 0 & 0.2309(2) & 2.27(30) &  & 0 & 0 & 0.2300(2) & 4.09(33) \\
O2 & 9b & 0.5397(1) & -0.5397(1) & 0.8503(1) & 3.32(18) &  & 0.5406(1) & -0.5406(1) & 0.8482(1) & 6.79(21) \\
O3 & 9b & 0.5620(1) & -0.5620(1) & 0.2723(1) & 6.01(21) &  & 0.5612(1) & -0.5612(1) & 0.2712(2) & 14.37(27)      \\ 
\end{tabular}
\label{table:11bm}
\end{table*}

\begin{figure*}[!t]  
\centering
\includegraphics[ height=3 in]{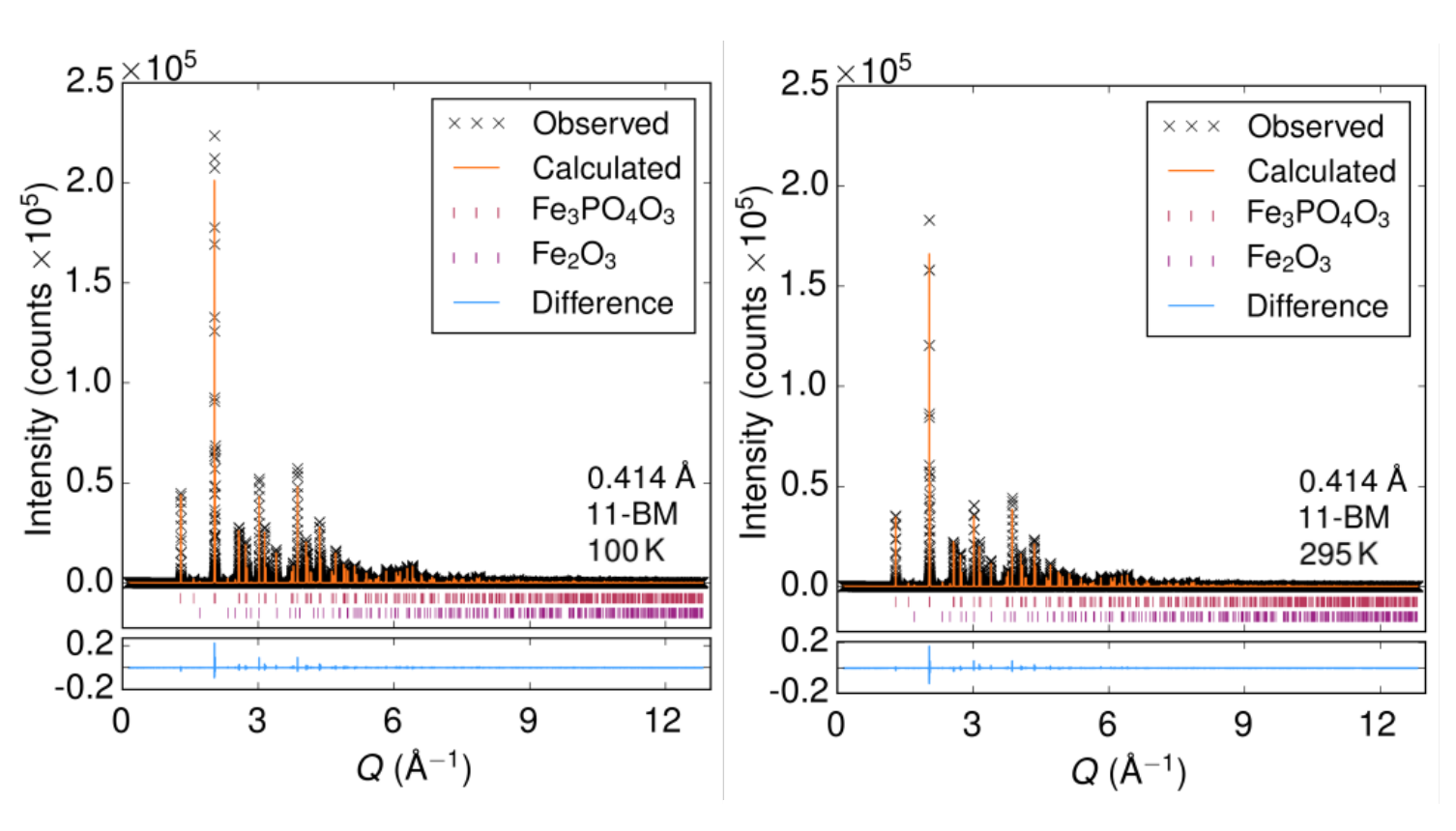}
\caption{High resolution X-ray diffraction Rietveld pattern refinements of \fepo \ at $T = 100\,$K (left) and $T = 295\,$K (right) collected at the Advanced Photon Source 11-BM-B beam line using an incident wavelength of 0.414 \AA. Observed data ($\times$) are plotted with the Rietveld refinement profile (orange line). A 3.56 wt\% Fe$_2$O$_3$ impurity is present in the sample (bottom ticks). No change in lattice symmetry is observed between 100 K and 295 K.} 
\label{fig:11bm}
\end{figure*}

\subsection{Synchrotron X-ray Diffraction on \fepo}

Structural refinements of SXRD data taken above (295 K) and below (100 K) the magnetic transition temperature were performed using the published crystal structure as a starting configuration.\cite{modaressi1983fe3po7}  No change in symmetry was observed on cooling through $T_N$.  A 3.56wt\% Fe$_2$O$_3$ impurity was included in the Rietveld refinements.    The SXRD data and refinements are shown in Fig. \ref{fig:11bm}.

The crystal structure of \fepo \ consists of three crystallographic sites for oxygen and unique sites for Fe and P.   Note that in the hexagonal setting of $R3m$ the overall fractional $z$ coordinate is not fixed by any of the special positions occupied by atoms in \fepo. We therefore chose an arbitrary fixed value of $z = 0.7425$ for Fe.   Table \ref{table:11bm} shows the refined positional and unit cell parameters, which display little variation from 295\,K to 100\,K, and the isotropic displacement parameters ($U_{iso}$) decrease upon cooling to 100 K, as expected.

\subsection{Neutron Diffraction}       
\label{sec:NPD}
\begin{figure*}[!htb]  
\centering
\includegraphics[ width=1.8\columnwidth]{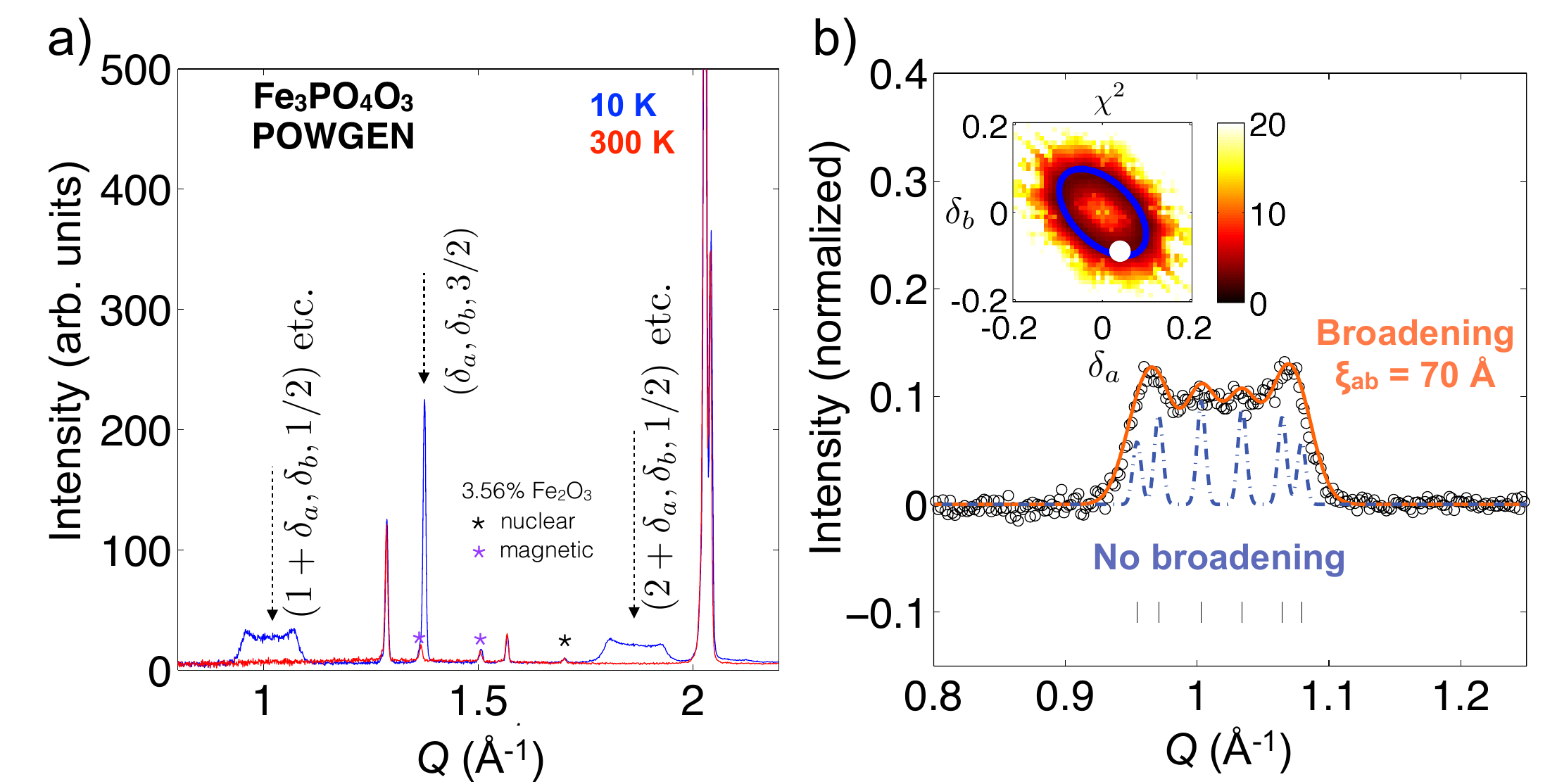}
\caption{ (a) Neutron powder diffraction from the POWGEN instrument at $T=10$ K  ($ <T_N$) and at $T= 300$ K ($>T_N$).  Broad magnetic peaks coexisting with a sharp peak appear below $T_N$.  The broad features span $\sim$ 18 times the instrumental resolution, while the sharp magnetic peak is resolution-limited.   The centers of all magnetic features can be indexed to a $\boldsymbol{k}_h = (0,0,3/2)$ ordering wavevector.  (b)  The width of the broad peaks can be partially accounted for by allowing an incommensurate modulation in the $ab$ plane, $\boldsymbol{k}_h = (\delta_a,\delta_b,3/2)$.  Gaussian broadening is applied to the lowest angle feature, which corresponds to several reflections that are nearly within the $a^*b^*$ plane; this additional broadening is necessary to reproduce the slope of the edges of the peak, implying short-range order with Gaussian correlations in the $ab$ plane with a correlation length of $\xi_{ab} = 70$ \AA.  Note that the free ion Fe$^{3+}$ magnetic form factor is not included in this comparison (see discussion in \ref{sec:magstruct}).  Including the form factor does not affect the necessary broadening (see Fig. \ref{fig:refinement} and related discussion).  Inset: the $(\delta_a,\delta_b)$ modulation used to generate the $(1+\delta_a, \delta_b, 1/2)$-type peaks shown in the main panel is represented by a white dot on a $\chi^2$ map showing that the best fits are obtained with a modulation wavevector of modulus $| \delta |$ =  0.073 \AA$^{-1}$ (blue circle).}  
\label{fig:peakshape}
\end{figure*}

\subsubsection{Magnetic Structure}
\label{sec:magstruct}
\fepo \ develops magnetic reflections upon cooling through  $T_N$ = 163 K.  Sharp (resolution limited) magnetic peaks coexist with broad, flat-topped peaks having widths approximately equal to eight (twenty) times the resolution of the BT1 (POWGEN) measurement.  The magnetic nature of these features is confirmed by comparison to the low temperature SXRD, which shows no extra peaks on cooling through the magnetic transition temperature (Fig. \ref{fig:11bm}).

The center of all magnetic features can be approximately indexed to a ${\boldsymbol k_{\text{h}}} = (0,0,1.5)$ ordering wavevector, where the ``$\text{h}$'' subscript refers to the hexagonal setting of the unit cell (in the rhombohedral setting, ${\boldsymbol k_{\text{r}}} = (0.5, 0.5, 0.5)$).  Such a wavevector produces a $\pi$ phase shift of the magnetic structure for each triangular layer along the hexagonal $c$ axis.  In order to reproduce the breadth of the flat-topped peaks, a modulation in the $ab$ plane must be added, i.e., $\boldsymbol\delta_{\text{h}} = (\delta_a,\delta_b, 0)$.   This splits reflections at zone centers that have nonzero components in the $ab$ plane, but does not split reflections having a purely $c$-axis component.   For instance, $\boldsymbol k_{\text{h}}$ = (0, 0, 1.5) becomes $(\delta_a, \delta_b, 1.5)$ and therefore only slightly shifts in $|Q|$ for small $\delta_a$ and $\delta_b$.  In contrast, reflections such as (1, 0, 0.5) and its 5 other symmetry-equivalent peaks become, e.g. $(1+\delta_a, \delta_b, 0.5)$,  $(\delta_a, 1+\delta_b, 0.5)$, $(1+\delta_a,-1+\delta_b, 0.5)$, etc., generating a series of six reflections that are closely-spaced in $Q$.

Inspection of the high resolution POWGEN data reveals that the shape of the broad peaks cannot be reproduced using a sum of closely spaced resolution-limited peaks (Fig. \ref{fig:peakshape}).  In fact, even a nearly continuous distribution of peaks, which could hypothetically result from a continuous degeneracy of in-plane modulation wavevectors, cannot account for the shape of these features (see the supplemental peakshape analysis in Fig. \ref{fig:peakshape2}).  This is evident from the slope of the edges of the broad features.   In order to account for the peak shape, some broadening of in-plane peaks must be applied.  This indicates finite sized domains in the $a$ and $b$ directions.  In order to reproduce the peak shape in the POWGEN data, we applied Gaussian broadening of FWHM = 0.03 \AA$^{-1}$ to the low angle reflections generated by a helical structure with $\boldsymbol k_{\text{h}} = (0.0259, -0.0902, 1.5)$ (the same magnetic structure refined from the BT1 data, discussed below).  This broadening is a factor of 6 greater than the instrumental resolution, and corresponds to an in-plane correlation length of only $\sim$ 70 \AA, while the helical modulation length is 86 \AA.  Importantly, however, the $c$ axis correlations remain long ranged, as indicated by the resolution-limited ($\delta_a$, $\delta_b$, 1.5) peak, implying correlations longer than 900 \AA.  Therefore, the diffraction pattern strongly suggests needle-like magnetic domains, with the longest correlations along the hexagonal $c$-axis, along which unmodulated antiferromagnetic correlations form.

The shape of broad peaks does not change on cooling to at least 4.5 K (3\% of $T_N$) (Fig. \ref{fig:BT1}).   This indicates that the short range magnetic domains are either equilibrium configurations or are kinetically inhibited from further growth.  Furthermore, the relevant wavevectors for the modulation do not change as a function of temperature.  No higher harmonics are observed at any temperature, which is consistent with the absence of the ``squaring up'' of a longitudinally modulated structure.\cite{jensen1991rare}  The possibility of a square-wave structure that develops immediately upon cooling through $T_N$ was also investigated and found to not reproduce the peak shape (Fig. \ref{fig:peakshape2}).   These considerations imply that the underlying magnetic structure is helical or conical with all ordered moments being of the same size.

In order to further refine the magnetic structure of \fepo, we used an ordering wavevector of the form ($\delta_a$, $\delta_b$, 1.5) in the Fullprof Rietveld refinement program.   Fullprof allows the refinement of anisotropic size broadening term with Lorentzian shape, accounting for either needle-like or platelet correlation volumes, for constant-wavelength NPD data. \cite{rodriquez2004line} We refined the Lorentizan broadening term for a needle-like correlation volume with long axis parallel to $c$, using the (constant-wavelength) BT1 data at $T=4.5$ K.  

The determination of the relative spin orientations proves difficult based solely on the powder diffraction data due to domain and powder averaging.  However, certain possibilities can be ruled out.  For instance, in order to generate the $(\delta_a, \delta_b, 1.5)$ peak (the observed sharp magnetic peak), there must be a net moment in the $ab$ plane in the structural unit cell.  This rules out helical variations of 120$^{\circ}-$type order on the triangular units.  The best fits can be obtained by starting with a collinear magnetic structure, with moments aligned on each triangle and a $\pi$ phase shift from layer to layer (i.e., a commensurate (0,0,1.5) parent structure), and then allowing an $ab$-plane modulation to produce a helical structure.  The best fit is obtained when the helical plane direction, $\hat{n}$, lies in the $ab$ plane; for example, the moments can be confined to the $ac$ plane as in the structure refinement shown in Fig. \ref{fig:refinement}.   Other types of phase-modulated structures cannot be ruled out, such as an antiferromagnetic conical structure with the cone axis along $a$ and the opening angle being $\sim 70^{\circ}$, though this type of structure would be less favored by Heisenberg exchange (Appendix \ref{sec:conical}).  We therefore choose the simplest structure, a helix as shown in Fig. \ref{fig:refinement}, for the purposes of further discussion.  In all of these types of structures (with the cone or helical axis in the $ab$ plane), the moment size refines to $\sim 4.1$ to 4.3 $\mu_B$, which is significantly lower than the 5 $\mu_B$ expected from the $S$=5/2 state of Fe$^{3+}$, but is consistent with a Curie-Weiss analysis of the magnetically dilute series \fexpo (see Section \ref{sec:CW}).  

In order to determine the direction of the \emph{modulation vector} in the $ab$ plane, $\boldsymbol{\delta}$, we systematically investigated a range of $(\delta_a, \delta_b)$ pairs by refining the following: the two angular parameters, $\theta$ and $\phi$, specifying the direction of the helical plane ($\hat{n}$) relative to the $c$ and $a$ axes, the magnetic moment, and the anisotropic (needle-like) Lorentzian anisotropic broadening, using the Fullprof refinement software in the helical magnetic structure mode.  The fitting range was restricted to $Q$ = 0.27 to 3.2 \AA$^{-1}$ in the BT1 data at $T = 4.5$ K.  The resulting $\chi^2$ values for the fits are shown in the $(\delta_a, \delta_b)$ plane in the inset of Fig. \ref{fig:peakshape}, which shows that only the \emph{magnitude} of the modulation wavevector can be constrained based on this fit.  The blue line shows the modulus of best fit of $|\delta| = 0.087 \pm 0.009$ r.l.u. \footnote{Note that since the ($\delta_a,\delta_b$) pair expresses components of the reciprocal lattice vector, with an angle of $60^{\circ}$ between the $a^*$ and $b^*$ directions, a line of constant modulus appears as an ellipse in such plots}   This shows that a particular modulation direction cannot be distinguished based on these data.  

Based on simple near neighbor interactions, we argue in Section \ref{sec:microscopic} that there may actually be no preferred direction for the modulation vector in the system, i.e., a single crystal diffraction pattern would hypothetically show incommensurate \emph{rings} of scattering.  On the line of constant modulus which produces the lowest $\chi^2$ values, $\theta$ does not deviate from 90$^{\circ}$ by more than 4$^{\circ}$, confirming that the helical plane contains the $c$ axis, while the magnetic moment does not deviate from 4.1 $\mu_B$ by more than 0.2 $\mu_B$, and the Lorentzian broadening term varies from 6.2 to 10.7, corresponding to correlation lengths of 45 to 85 \AA.  Visual inspection of various choices for $\boldsymbol{\delta}$ along this line of ``best fit'' favors a fit with $\boldsymbol{\delta} = (0.028, -0.097)$, shown in Fig. \ref{fig:refinement}, and here the broadening term is 6.9, corresponding to $\xi_{ab}  \sim 70$ \AA, in agreement with our peak shape analysis of the POWGEN data using same magnetic structure (Fig. \ref{fig:peakshape}).

Finally, we note that the lowest angle broad peak is observed to be rather symmetric in $Q$, whereas the model, which includes the spherical form factor for Fe$^{3+}$ as parameterized in Ref. \onlinecite{brown2006magnetic}, does not fully reproduce this.  We made many attempts to find a magnetic structure that would produce the symmetric shape of the lowest angle feature, while retaining other important features, but these attempts were not fruitful.  We note that removing the form factor allows a better fit of the lowest angle feature (see Fig. \ref{fig:peakshape}). Although it is not physical to completely neglect the form factor, a flatter form factor near this low angle peak can arise from covalent spin density.  Similar effects have been studied in detail for other transition metal ions; for instance, Ni$^{2+}$ is known to have a more contracted spin density in the compound NiO compared to the free ion, leading to a more gradual decrease in the form factor.\cite{owen1966covalent}   Given the mounting evidence for ligand charge (and spin) transfer in \fepo (see Section \ref{sec:reducedmom}), the moment reduction and poor agreement with conventional magnetic form factors are therefore potentially related. Further measurements on single crystal samples will be required to understand details of the magnetic form factor, the domain anisotropy, and the spin orientations.

\subsubsection{Temperature Dependence from Neutron Powder Diffraction}

\begin{figure*}[!htb]  
\centering
\includegraphics[ width=1.8\columnwidth]{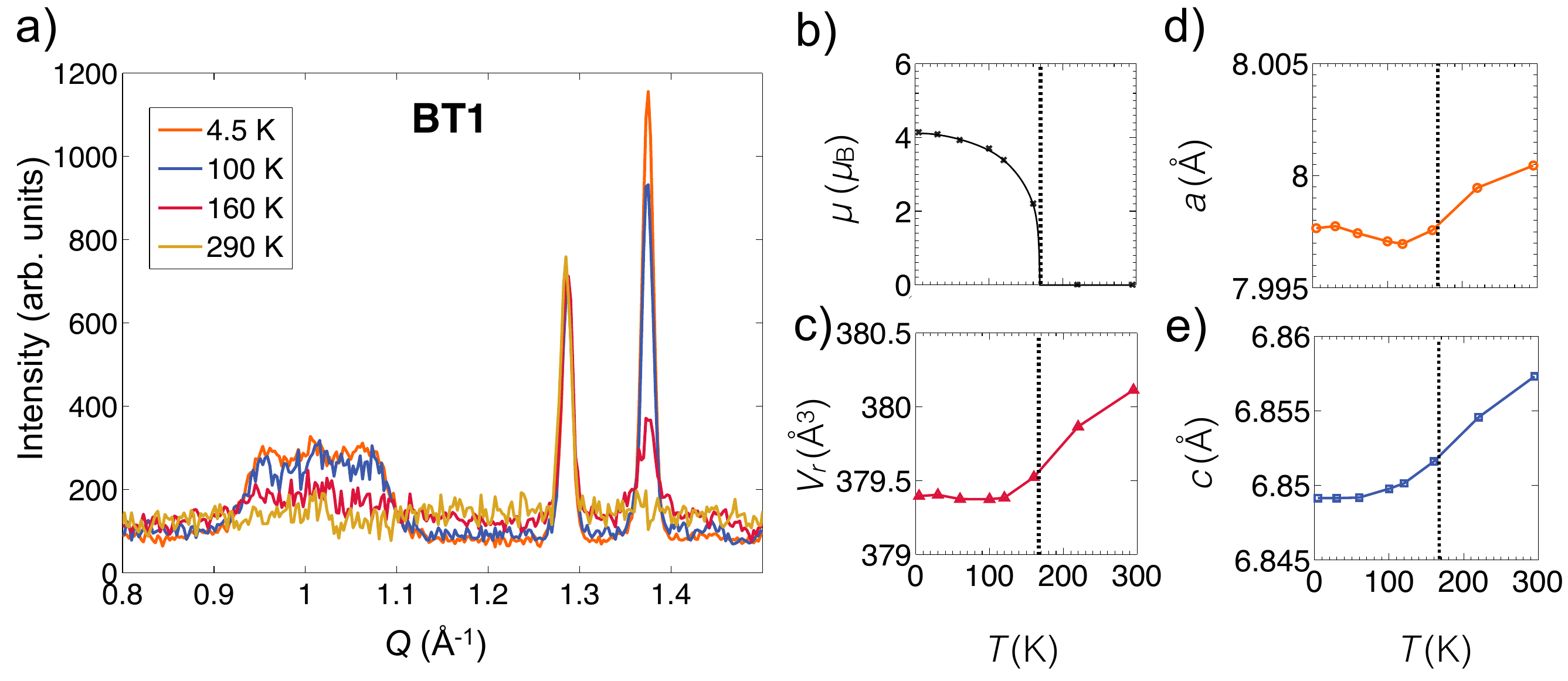}
\caption{Temperature dependent results of NPD from \fepo \ taken on the BT1 instrument.  (a)  Evolution of the diffraction profile near the lowest angle magnetic peaks.  (b) Magnetic moment, $\mu$, vs. $T$ from the magnetic structure refinement shown in Fig. \ref{fig:refinement}.  Solid line is a guide to the eye.  (c) Unit cell volume in the hexagonal setting; a small negative thermal expansion is observed below 100 K.  d) and e) Temperature dependence of the unit cell lengths $a$ and $c$, respectively.  The cell length, $a$, decreases on cooling but begins to increase below $T_N$. Errorbars are smaller than markers in b) through e).}
\label{fig:BT1}
\end{figure*}

\begin{figure}[!tb]  
\centering
\includegraphics[ width=1\columnwidth]{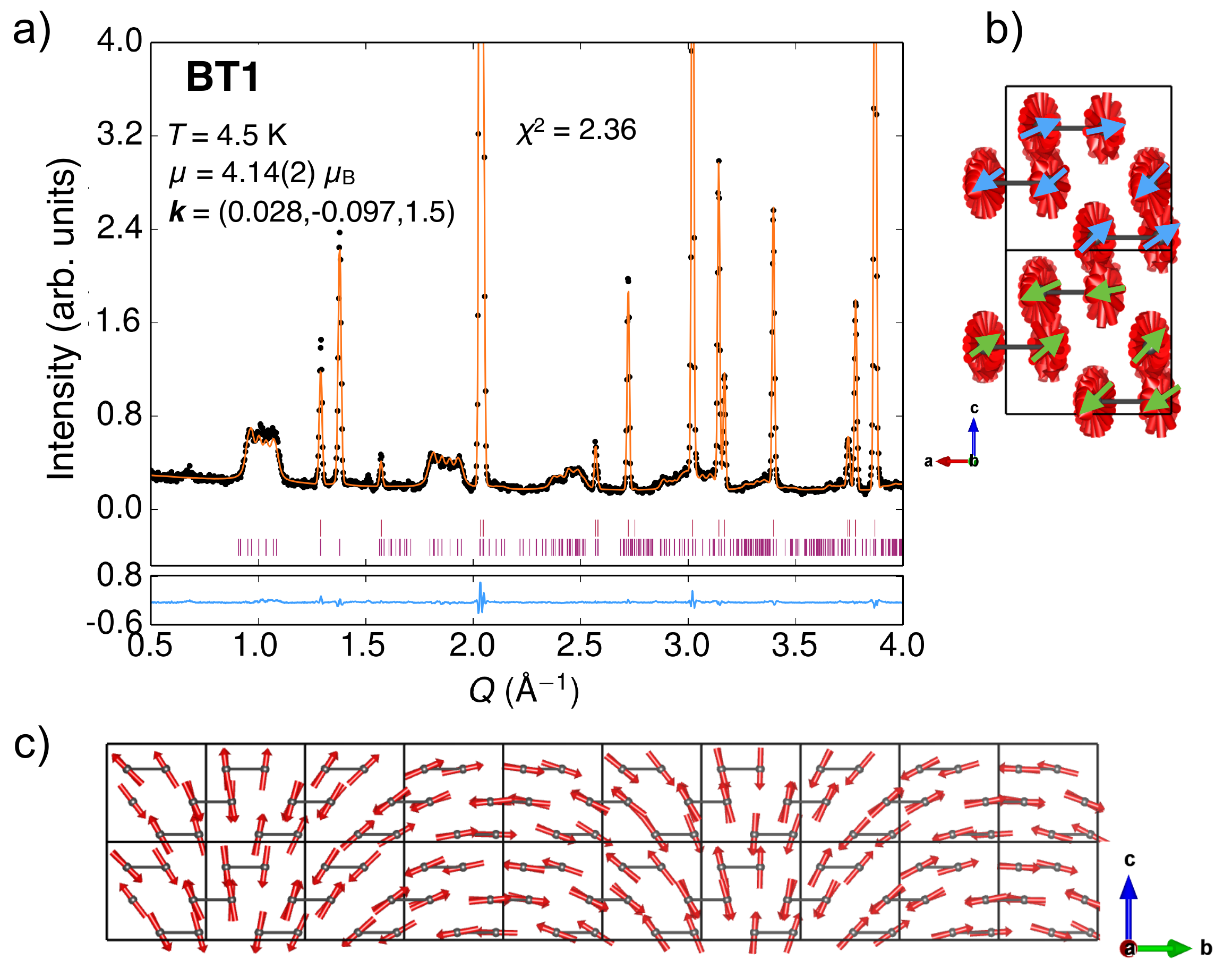}
\caption{ (a) Results of Rietveld refinement of the BT1 neutron powder diffraction data at 4.5 K.  The nuclear (top ticks) and magnetic (bottom ticks) phases are shown.  Broadening of in-plane magnetic peaks was applied (see text).  The difference curve is shown in the bottom box (blue line).  (b) Illustration of of the refined helical magnetic structure, showing 10 unit cells superimposed along $b$.  The highlighted arrows show the moment directions in the nearest unit cells.  Thick black lines connecting atoms are $J_1$ bonds (i.e., the edges of the triangular units). (c) View of the magnetic structure along the $a$ axis, which is perpendicular to the helical plane.  Note in b) and c) how moment directions are reversed in neighboring unit cells along the $c$-axis. }
\label{fig:refinement}
\end{figure}

The thermal dependence of the basic structural parameters were determined using the BT1 data, by refining a nuclear and magnetic phase at each temperature (taking the planar helix structure shown in Fig. \ref{fig:refinement} as the magnetic structure).  The lattice parameters (unit cell lengths $a$ and $c$), cell volume, and ordered magnetic moment are shown as function of temperature in Fig. \ref{fig:BT1} b). The $a$ unit cell length and the cell volume vary non-monotonically below $T_N = 163$ K; $a$ begins to \emph{increase} below $\sim 100$ K, exhibiting a minimum just below $T_N$.   Meanwhile, $c$ does not decrease enough to compensate the increase in $a$; a corresponding negative thermal expansion is observed below 100 K.  These trends may be correlated with the onset of magnetic order, but the current measurements are not conclusive on this point.  The other structural parameters, such as anisotropic displacement parameters, cannot be refined accurately from this data, as the experiment was optimized for magnetic (low $Q$) scattering.

\section{\label{sec:level1}Discussion}

\subsection{Curie-Weiss analysis and reduced Fe$^{3+}$ moment}
\label{sec:reducedmom}

Previous Curie-Weiss analysis of the magnetic susceptibility in \fepo \ found effective magnetic moments of $p_{\text{eff}} = 6.45 \mu_B$ \cite{modaressi1983fe3po7} and  $4.3 \mu_B$,\cite{gavoille1987magnetic} and Curie-Weiss temperatures of  -1707 K \cite{modaressi1983fe3po7} and -1087 K.\cite{gavoille1987magnetic}  With five electrons in five $3d$-orbitals, Fe$^{3+}$ most often takes on a spin-only $S$=5/2 angular momentum, with an expected effective moment of $p_{\text{eff}} = 5.9 \mu_B$.  Therefore the magnetic moments determined from the previous Curie-Weiss analysis are anomalous and mutually inconsistent.  Examination of the inverse susceptibility in the temperature ranges used for fitting reveals why this is the case; the curve is non-linear up to at least 900 K.  The obtained Curie-Weiss temperatures, $\Theta_{CW}>$1000 K, give a hint as to why this may be: the temperature range used for fitting is exceeded by the deduced mean field interaction strength, invalidating the mean field approximation.   Strong magnetic correlations above $T_N$ are also revealed from the NPD patterns (Fig. \ref{fig:BT1}); significant diffuse scattering around the (0, 0, 1.5) position is evident above 160 K.  This is expected in frustrated magnets.

By this reasoning, the true mean field interaction strength, normally encoded by $\Theta_{CW}$, is presently unspecified, but must be greater than $\sim$ 900 K.   The effective moment can still be deduced by diluting the magnetic lattice, as we have done using the \fexpo \ series (Fig. \ref{fig:CW}).  The Curie-Weiss analysis of that series shows that $\Theta_{CW}$ decreases with increasing $x$ (as expected), reaching a value that falls within our fitting range (300 K - 700 K) around $x = 1.5$.  The effective moment per Fe$^{3+}$ approaches $p_\text{eff} \sim$ 5.1 $\mu_B$ at these higher $x$ values.   

What could be the cause of this moment reduction?  Orbital contributions are not expected in the half filled shell of Fe$^{3+}$, so the $g$-factor is not likely to be significantly reduced.  The most likely possibility is ligand charge transfer, i.e., 2p(O$^{2-}$) $\rightarrow$ 3d(Fe$^{3+}$), which reduces the percentage of unpaired spins on Fe.   Such effects are not uncommon and have been proposed for other Fe$^{3+}-$containing compounds such as $\alpha$-Fe$_2$O$_3$ (Hematite), \cite{hill2008neutron} and Ca$_3$Fe$_2$Ge$_3$O$_{12}$.\cite{plakhty1999spin} Indeed, such covalency effects were previously proposed for \fepo.\cite{gavoille1987magnetic}

The reduced effective moment is consistent with a $g$-factor of 2 and a reduced total spin value of $S$ = 2.1 per Fe.  This leads to a microscopic moment of $\mu = 4.2 \mu_B$, which is consistent with our refined value of the ordered magnetic moment from NPD (4.14(2) $\mu_B$), recalling that quantum fluctuations in antiferromagnets tend to further reduce the \emph{ordered} moment.  Thus, we conclude that the moment associated with Fe in \fepo \ is reduced by 16\% compared to the expected $\mu = 5 \mu_B$ for an isolated Fe$^{3+}$ ion.

\subsection{Microscopic Origin of Helimagnetism}
\label{sec:microscopic}

Two common sources for helimagnetism are 1) the anti-symmetric Dzyaloshinskii-Moriya (DM) interaction, and 2) frustration arising from a competition between $J_1$ and $J_2$.   We first discuss the effect of DM interactions in \fepo. 

In \fepo, DM interactions are allowed for each pair of Fe ions.  Here we consider only nearest neighbors.  A mirror plane bisects each nearest neighbor bond.   According to symmetry arguments first articulated by Moriya,\cite{moriya1960anisotropic} this leads to a DM vector constrained to lie in the plane perpendicular to the $J_1$ bond. The triangular geometry dictates that any component of the DM vector lying in the $ab$ plane for a given near neighbor bond will then be cancelled by the remaining two, leaving only the possibility of a \emph{net} DM vector along the $c$-axis.  Therefore the nearest neighbor DM interaction could only favor spin structures having a net pairwise cross product pointing along $\pm (0,0,1)$. We made many attempts to find a suitable spin structure based on this constraint, but neither a helical nor conical structure with a net cross product along $c$ produces a satisfactory fit to the data.    In fact, the best fits are obtained from configurations with a net cross product \emph{perpendicular} to $c$.

The most likely cause for the observed helimagnetism is therefore the competition between AFM $J_1$ and $J_2$.   The type of magnetic order that develops in \fepo \ indicates a dominant antiferromagnetic $J_2$.  Since $J_2$ couples triangular units in different $c$-axis layers, it is responsible for the $\pi$ phase shift between each triangular layer.    Meanwhile, the nearest neighbor exchange within the triangles may be either ferromagnetic or antiferromagnetic.   Modaressi, \emph{et al.} analyzed the possible magnetic couplings, both direct and indirect, and proposed both ferromagnetic and antiferromagnetic contributions for $J_1$ and only antiferromagnetic coupling for $J_2$.\cite{modaressi1983fe3po7}  This may allow for a weaker overall $J_1$ compared to $J_2$ (both AFM).  Furthermore, since the coordination number for $J_2$ bonds is $z=4$ while that for $J_1$ bonds is $z=2$, comparable AFM $J_1$ and $J_2$ strengths would still produce a state dominated by $J_2$.  For example, they could produce a commensurate version of the proposed helical state, with all spins aligned on a triangle and a $\pi$ phase shift from layer to layer (this is symbolically represented in Fig. \ref{fig:medial} b).  We used the software package SpinW \cite{spinW} to numerically minimize the energy of the AFM $J_1 -J_2$ Heisenberg model for \fepo, starting from random spin configurations in a 10$\times$10$\times$4 supercell with periodic boundary conditions.  We found that the aforementioned ``parent'' commensurate structure is realized for the ratios $J_2/J_1 > 3.5$, and a helical incommensurate structure like the one refined for \fepo \ is realized for $J_2/J_1 \sim 3 $.   The helical modulation therefore arises as a compromise for the competing interactions; the commensurate structure internally frustrates the triangular units and this can be relieved by slightly canting spins away from perfect co-alignment.  The pitch of the helical modulation, i.e., $|\delta|$, should then be controlled by the ratio of $J_1$ to $J_2$.  In \fepo \ we found the pitch to be 86 \AA; this number can in principle be used to extract the ratio of exchange parameters using a more detailed calculation. Considering only these completely isotropic exchange interactions, the triangular geometry prevents this mechanism from producing a preferred modulation vector direction in the $ab$ plane.  In principle, this could lead to domains forming from unique selections from a continuous distribution of $\boldsymbol{\delta}$ directions, which could contribute to the apparent independence of $\chi^2$ for the magnetic refinements on the direction of the modulation vector, as shown in Fig. \ref{fig:peakshape}.  However, it is more likely that small anisotropies in the interactions, or even entropic terms in the free energy, could select preferred directions (for instance, leaving a three-fold degeneracy).  Such information would likely be obscured due to powder averaging the neutron diffraction pattern. The simple arguments presented here also do not offer a motivation for the modulation vector, $\boldsymbol{\delta}$, to be locked to the helical plane direction, $\hat{n}$, in contrast to MnSi.\cite{plumer1981wavevector}  To first order, these two directions ($\boldsymbol{\delta}$ and $\hat{n}$) may be independent  in \fepo.

\subsection{Highly Anisotropic Domains}

 \begin{figure}[!tb]  
\centering
\includegraphics[ width=1\columnwidth]{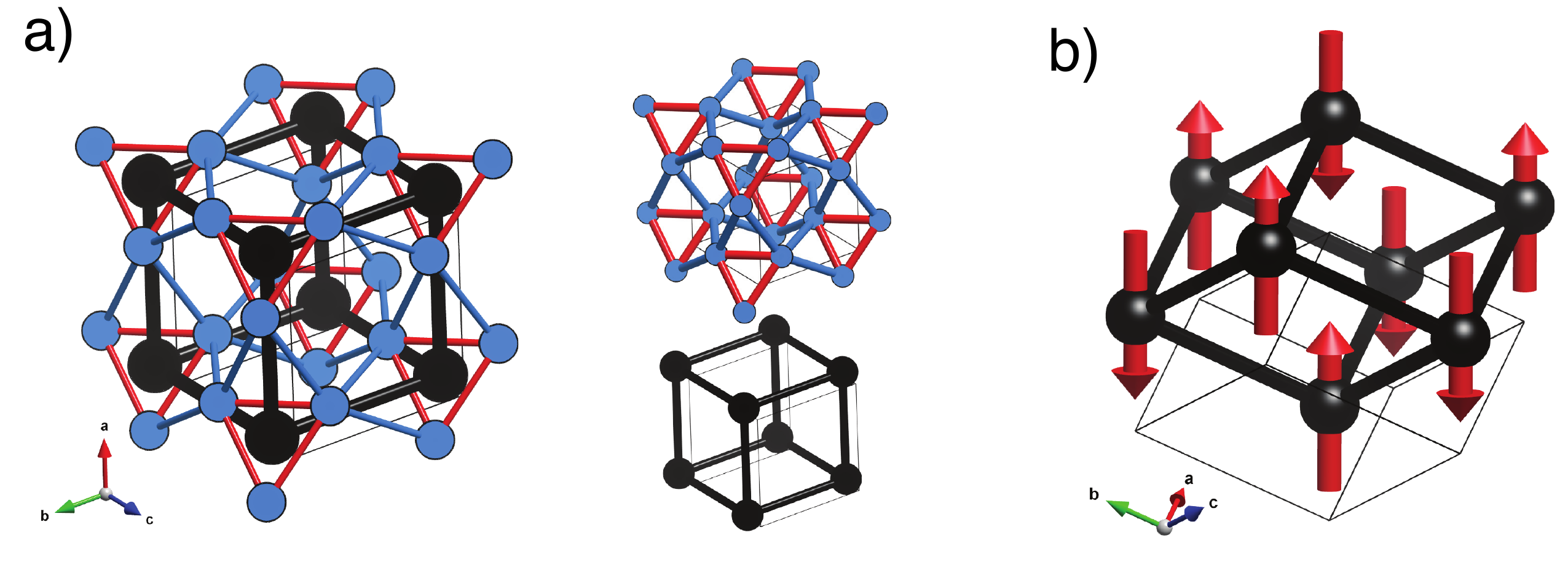}
\caption{ Premedial lattice in \fepo. Unit cells represented in rhombohedral setting. (a) Top right: schematic of iron sites in \fepo \ connected by red ($J_1$) and blue ($J_2$) bonds. Bottom right: locations of the centers of triangles formed by the $J_1$ bonds, i.e., the simple rhombohedral ``premedial'' lattice.  Left: the relative positions of the premedial lattice points (black) with the iron sites (blue). (b) abstraction of the type of \emph{commensurate} magnetic structure that the helical state in \fepo \ is based upon.  Each triangular unit possesses nearly collinear aligned moments; this direction is represented by a single arrow on the premedial lattice, and forms a $G$-type antiferromagnet on a simple rhombohedral lattice.  }
\label{fig:medial}
\end{figure}

Perhaps the most interesting feature of the magnetic structure adopted by \fepo \ is the needle-like AFM domains that maintain small correlation lengths (70 \AA) in the $ab$ plane, but extended correlations ($>900$ \AA) along $c$.   The needle-shaped correlation volume implies that domain walls cost little energy as long as they are ``vertical'', i.e., parallel to the hexagonal $c$ axis.  

These anisotropic domains are counterintuitive, given that the interactions are three-dimensional.  As noted in the previous section, the magnetic structure implies that $J_2$ is strong and antiferromagnetic.  $J_2$ connects the Fe$^{3+}$ triangular units three dimensionally.  This can be visualized using the ``premedial lattice'', in which a lattice point is placed at the center of each triangle.\cite{henley2010coulomb}  The premedial lattice for \fepo \ is a simple rhombohedral lattice; i.e., a simple cubic lattice compressed along a (111) direction (Fig. \ref{fig:medial} a). Each bond between nearest neighbors on the premedial lattice represents two $J_2$ bonds in the normal lattice.  One can visualize the commensurate version of the magnetic structure realized in \fepo \ by assigning a single arrow for each triangular subunit (premedial lattice point).  The result is a G-type antiferromagnet on a distorted simple cubic lattice, a canonical three dimensional antiferromagnet (Fig. \ref{fig:medial} b).

\begin{figure}[!tb]  
\centering
\includegraphics[ width=\columnwidth]{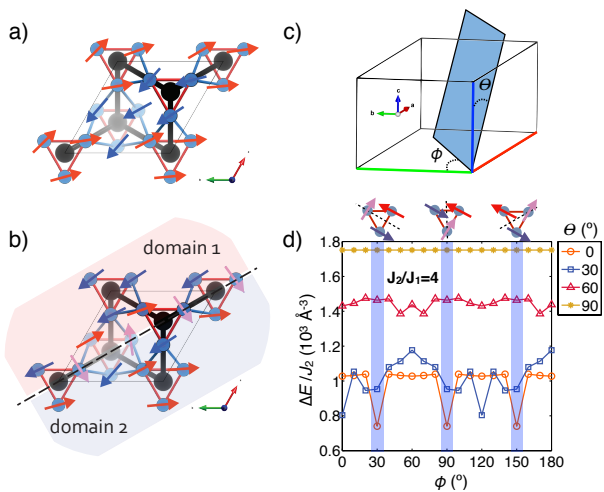}
\caption{ Understanding domain walls parallel to $c$ in \fepo.  (a) Cartoon showing a helical incommensurate structure similar to the one found in \fepo which can be generated from competing AFM $J_1$ and $J_2$, with $J_2>J_1$.  (b)  A simple 180$^{\circ}$ domain wall passing through the mirror planes of the $J_1$ triangular units; spins lying on the domain wall are rotated by $\sim 90 ^{\circ}$ from the local magnetization direction.  (c) Definition of angular parameters for the orientation of a domain wall.  (d) Simulated energy cost for a domain wall with orientation angles $\phi$ and $\theta$ as shown in c), for $J_2$/$J_1$ = 4.  The lowest energy cost is for ``vertical'' (parallel to $c$) domain walls passing through local triangular mirror planes ($\theta = 0$, $\phi = 30, 90, 150 ^{\circ}$). }
\label{fig:domain}
\end{figure}

Given these strong three dimensional couplings, why does \fepo \ adopt such highly anisotropic correlations?  The appearance of vertical domain walls in this structure can be partially understood as a consequence of the frustrated triangular units that decorate the premedial lattice.  The frustration arising from $J_1$ - $J_2$ competition on the normal lattice allows this simple 3D structure to be fractured by domain walls since they are favored by the $J_1$ AFM interaction.  Consider a simple domain wall consisting of a 180$^{\circ}$ spin reorientation at a plane.  To investigate the most favorable orientation for such a domain wall, we have considered the energy cost for a 180$^{\circ}$ domain wall in the \emph{commensurate} parent magnetic structure by assuming that when the domain wall plane intersects a lattice site, the moment on that site lies perpendicular to the (arbitrary) direction chosen for the collinear spin structure (Fig. \ref{fig:domain} a and b).   The orientation of the domain wall is parameterized by two angles, $\theta$ and $\phi$, measured from the hexagonal $c$ and $b$ axes, respectively (Fig. \ref{fig:domain} c).  For $J_2$/$J_1$ = 4, i.e., within the parameter range found to produce the parent commensurate structure of interest, the energy cost per $J_2$ per unit volume is shown in Fig. \ref{fig:domain} d) as a function of $\phi$, at different $\theta$ values.  In general, the lowest energies are achieved at low $\theta$ values, corresponding to ``vertical'' domain walls.  Furthermore, the minimum energy is along the local mirror planes for the triangular units, as illustrated at the top of Fig. \ref{fig:domain} d).  Therefore, for the simple domain walls considered here, a threefold degeneracy of orientations exists.  The intersection of three such domain walls could occur at a point, producing a line defect.  For more realistic domain walls, these special points could be topological in nature, similar conceptually to the discrete vortex topological defects found in the hexagonal manganites.\cite{artyukhin2014landau}

Although helical magnetic structures are common in frustrated systems, \fepo \ is unusual in that it retains only short range correlations at temperatures 40 times below the ordering transition.  In many cases an intermediate incommensurate structure will be entropically stabilized above a transition to long range commensurate order, a so-called ``lock-in transition''.  An illustrative example is that of LiNiPO$_4$.  In this orthorhombic system, an intermediate helical incommensurate structure with short range correlations, similar to that found in \fepo, is formed in a small temperature range above $T_N$ = 19 K, at which a commensurate AFM structure is realized. \cite{vaknin1999weakly,vaknin2004commensurate} Preceding the transition to commensurate order, the incommensurate wavevector (modulation length) shrinks as the temperature is lowered, before ``locking in'' to zero at $T_N$.  Higher harmonic peaks are observed to grow in as the lock-in transition is approached, and unusual anisotropic AFM domains are observed in this low temperature phase.\cite{van2008anisotropy}  In contrast, the short range incommensurate structure in \fepo \ seems to be stable at all temperatures $T < 163$ K.  There is no change in the incommensurate modulation ($|\delta|$) as a function of temperature, from 160 K down to 4.5 K, and no higher harmonics have been observed.  Furthermore, no increase in the magnetic correlation length is observed over this temperature range.  Therefore, the short range incommensurate structure in \fepo \ is stabilized, either thermodynamically or perhaps kinetically (if the domain walls are pinned at lattice defects, for example).

In general, the equilibrium stabilization of AFM domain walls is not well-understood.   In ferromagnets, domain walls are stabilized by demagnetization fields associated with the net magnetic moment of each domain.  However, in an ideal antiferromagnet there is no such long-range potential to drive fractionalization.   A possible explanation, which so far has been the main route for theoretical exploration of this problem for general antiferromagnets, is the build up of magnetoelastic strain fields.\cite{minakov1990magnetostriction, gomonay2002magnetostriction}  In \fepo \ such strain fields could be the source of the small negative thermal expansion we have observed below 100 K (Fig. \ref{fig:BT1}).  The extent of magnetoelastic coupling in \fepo \ may be important to explore further in order to explain the stabilization of such a high density of AFM domain walls.  

Interest in AFM domains and topological defects is rapidly growing due to their importance in exchange bias effects and with the new proposals for generation of antiferromagnetic Skyrmions.  However, they are difficult to directly investigate experimentally due to the net zero moment.   Several techniques have been successfully employed in the past in order to observe AFM domains or their dynamics, for instance, polarized neutron tomography,\cite{schlenker1978neutron} optical second harmonic generation,\cite{van2008anisotropy} and photocorrelation spectroscopy.\cite{shpyrko2007direct, chen2013jamming}   Given its unusually high domain wall density and the potential for generating topological defects, it would be of great interest to explore the domain pattern in \fepo \ using these techniques.  The development of single crystal samples will be of utmost importance for this effort, as well as for future neutron scattering studies on this unusual material.

\section{\label{sec:level1}Conclusions}

We have investigated the magnetic and structural properties of \fepo \ using thermodynamic probes, synchrotron X-ray diffraction, and neutron powder diffraction.  We find that the coexistence of broad, flat-topped magnetic peaks with sharp magnetic peaks, as observed by neutron powder diffraction, is best described by an antiferromagnetic incommensurate helical structure with highly anisotropic domains that develops below $T_N=163$ K.  The domains are small within the hexagonal $ab$ plane, extending to only 70 \AA, while extending to at least 900 \AA \ along the hexagonal $c$-axis, implying a high density of \emph{``vertical''} domain walls ($\rho_{DW}$ = 0.65 nm$^{-2}$).  Stabilization of these antiferromagnetic domain walls may occur through magnetoelastic strain;  our temperature dependent structure refinements indicate negative thermal expansion below $T_N$.

The magnetic moment determined from our magnetic structure refinements, $\mu = 4.14(2) \mu_B$, is consistent with the effective moment determined from paramagnetic susceptibility on the magnetically dilute \fexpo \ series; $p_\text{eff} / \text{Fe} \rightarrow 5.1 \mu_B$ as the high temperature susceptibility reaches the truly paramagnetic regime.  This 16\% reduced moment compared to the $S$=5/2 expected for Fe$^{3+}$ could be an indication of covalency (ligand charge transfer) in \fepo.   

  The type of magnetic structure formed by \fepo \ indicates frustration between the antiferromagnetic exchange couplings, $J_1$ and $J_2$, with dominant $J_2$.  We have shown that despite the three dimensional network formed by these strongest interactions in \fepo, the presence of frustrated triangular $J_1$ subunits can lead to anisotropic needle-like antiferromagnetic domains, as observed.  The frustrated triangular subunits provide a natural location for fracturing the long range magnetic structure, and the ``path of least resistance'' is to propagate these defects along the $c$-axis.

Given the helical antiferromagnetic nature of the magnetic structure, as well as its propensity to form defects (domain walls), \fepo \ may be of interest to study in the context of topological spin textures such as antiferromagnetic Skyrmions. \cite{zhang2015antiferromagnetic}   Of more fundamental importance is the presently poorly-understood mechanism for stabilization of domains in antiferromagnets.  In this context, \fepo \ provides an unusual test case in which a high density of small domains form and persist to the lowest temperatures, which remains unexplained from either an equilibrium or kinetic standpoint.  For these reasons, the magnetic domain structure in \fepo \ certainly deserves further investigation, particularly in single crystal samples, should they become available.

\acknowledgements
We acknowledge the support of the National Institute of Standards and Technology, U. S. Department of Commerce, as well as Oak Ridge National Laboratory, U. S. Department of Energy, in providing the neutron research facilities used in this work.  We acknowledge the support of Argonne National Laboratory, U. S. Department of Energy, in providing the synchrotron facility used in this work.  We thank C.M. Brown and A. Huq for technical assistance with the neutron powder diffraction experiments.   KAR was supported by NSERC of Canada.

\newpage
\appendix
\section{\fexpo\ solid solution x-ray data}
\label{sec:fexpoxray}

\begin{figure}[!htb]  
\centering
\includegraphics[ width=\columnwidth]{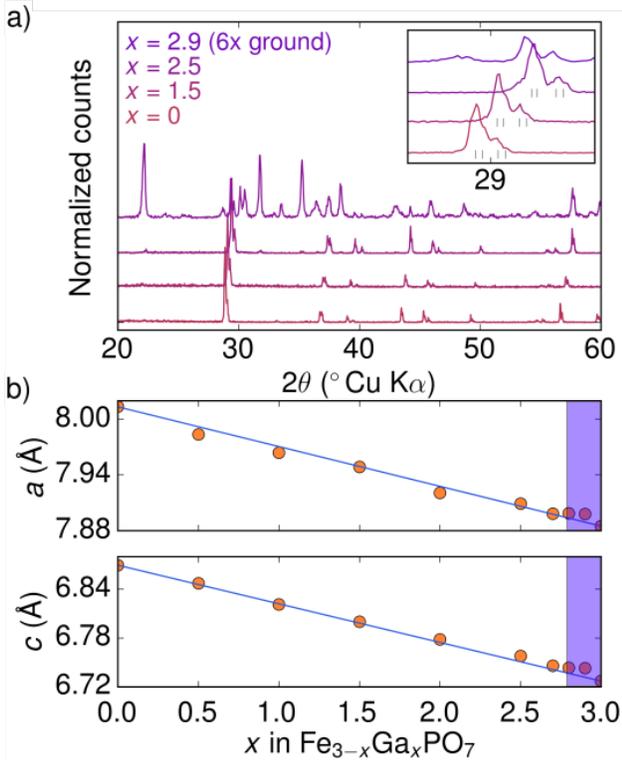}
\caption{ X-ray analysis of the \fexpo \ series.  (a) X-ray diffraction patterns (Cu-K$\alpha$ radiation) of select members of the series, showing the progressive increase of scattering angles for equivalent peaks (i.e., decrease in lattice parameters) with increasing Ga content ($x$).  Between $x=2.8$ and 2.9, the \fepo \ structure type cannot be produced as the majority phase, despite many regrindings.  (b) Linear trends in $a$ vs. $x$ and $c$ vs. $x$, as expected based on Vegard's law (errorbars are smaller than the data points).  Blue shaded region shows the range over which the desired structure cannot be produced as the majority phase (solubility limit reached).  The $x=3.0$ values are from a high pressure synthesis of Ga$_3$PO$_4$O$_3$ from Ref. \onlinecite{boudin1998ga3po7}.}
\label{fig:fexpoxray}
\end{figure}

The solid solution series \fexpo \ was investigated using a Scintag X-ray powder diffractometer (Cu-K$\alpha$ radiation).  Select diffraction patterns are shown in Fig. \ref{fig:fexpoxray} along with the refined lattice parameters as a function of $x$.   These follow a linear trend, as expected for solid solutions.  The solubility limit is reached between $x = 2.8$ and $x = 2.9$, where the product phase-separates into GaPO$_4$ and Ga$_2$O$_3$ with a small percentage (1.95\% by mole) of a phase with the \fexpo \ structure type.

 \begin{figure}[!t] 
\centering
\includegraphics[ width=\columnwidth]{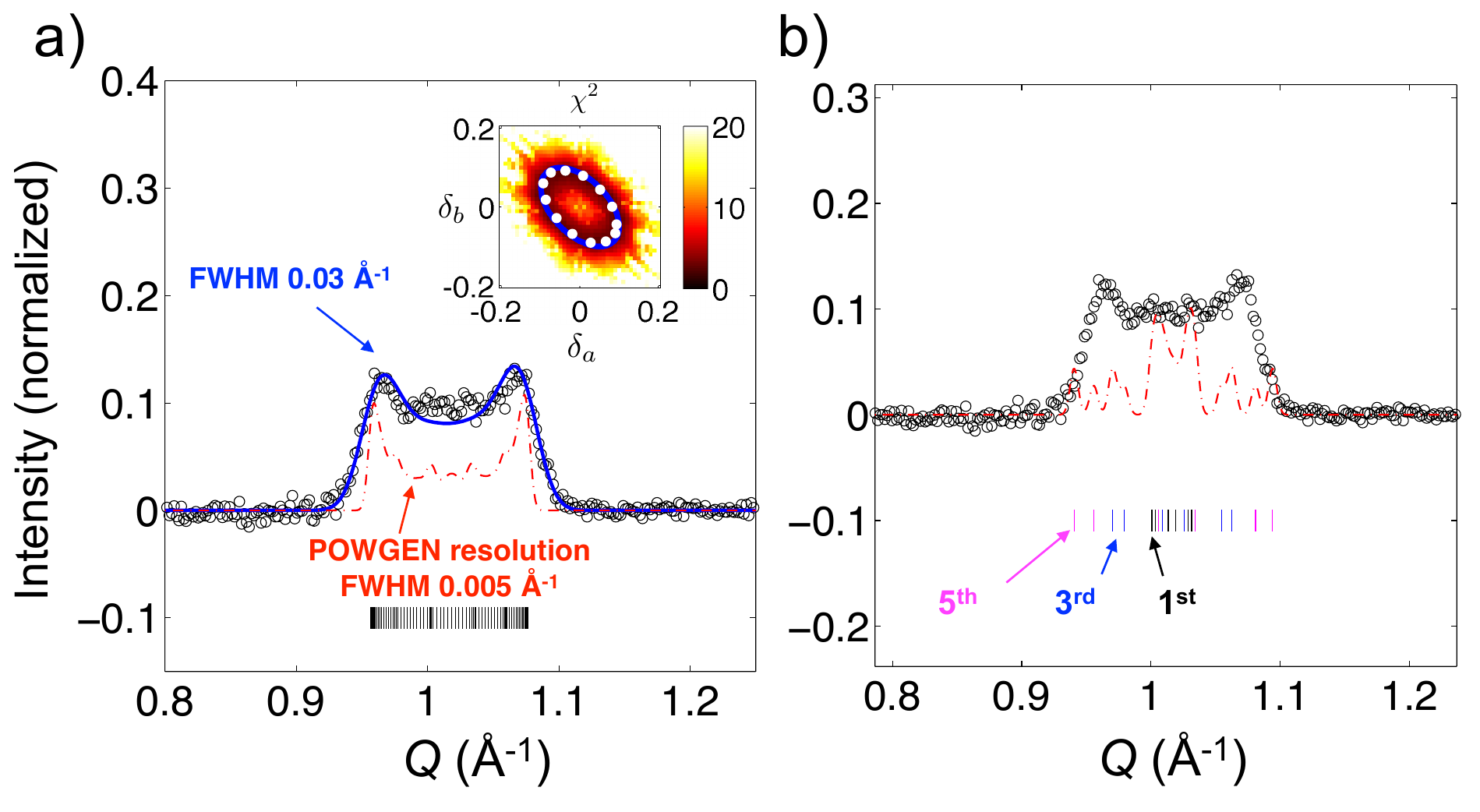}
\caption{Comparison of the peak shape for the lowest angle broad feature as observed with the POWGEN instrument to two models. (a) The same helical model as proposed for \fepo \ (Fig. \ref{fig:refinement}), but combining a nearly continuous distribution of wavevectors.  (b) Square wave model with up to the fifth harmonic shown (black, blue and magenta ticks show locations of $1^{st}$, $3^{rd}$ and $5^{th}$ harmonic peaks respectively).}
\label{fig:peakshape2}
\end{figure}

\section{Further details of neutron powder diffraction peak shape}
\label{sec:peakshape_supp}

The peakshape of the lowest $Q$ feature in the POWGEN NPD pattern for \fepo \ was further investigated using two additional models.  The first is the same helical model described in the main text, but with several choices for the modulation vector superimposed (see inset of Fig. \ref{fig:peakshape2} for sampling of ($\delta_a$, $\delta_b$, 1.5) wavevectors).  The result is a nearly continuous distribution of peaks within the broad feature (Fig. \ref{fig:peakshape2} a).  This model still requires a Gaussian peak broadening of FWHM = 0.03 \AA$^{-1}$ to account for the slope of the edges of the broad feature.  Note that no magnetic form factor is applied to the calculated intensities.  

The second model investigated is an incommensurate square wave model.  The square wave would produce higher harmonics (i.e. $k_n = (n \delta_a, n\delta_b, 1.5)$, with $n$ odd) in the Fourier expansion of the periodic structure, with intensities reduced by a factor of $1/n$ for the $n^{th}$ harmonic peaks.  Fig. \ref{fig:peakshape2} b) shows how a hypothetical square wave structure compares to the measured peakshape uisng $k_1 = (0.015, 0.008, 1.5)$, including the $1^{st}$, $3^{rd}$ and $5^{th}$ harmonics.

\section{Alternative Conical Magnetic Structure for \fepo}
\label{sec:conical}
\begin{figure}[!thb]  
\centering
\includegraphics[ width=\columnwidth]{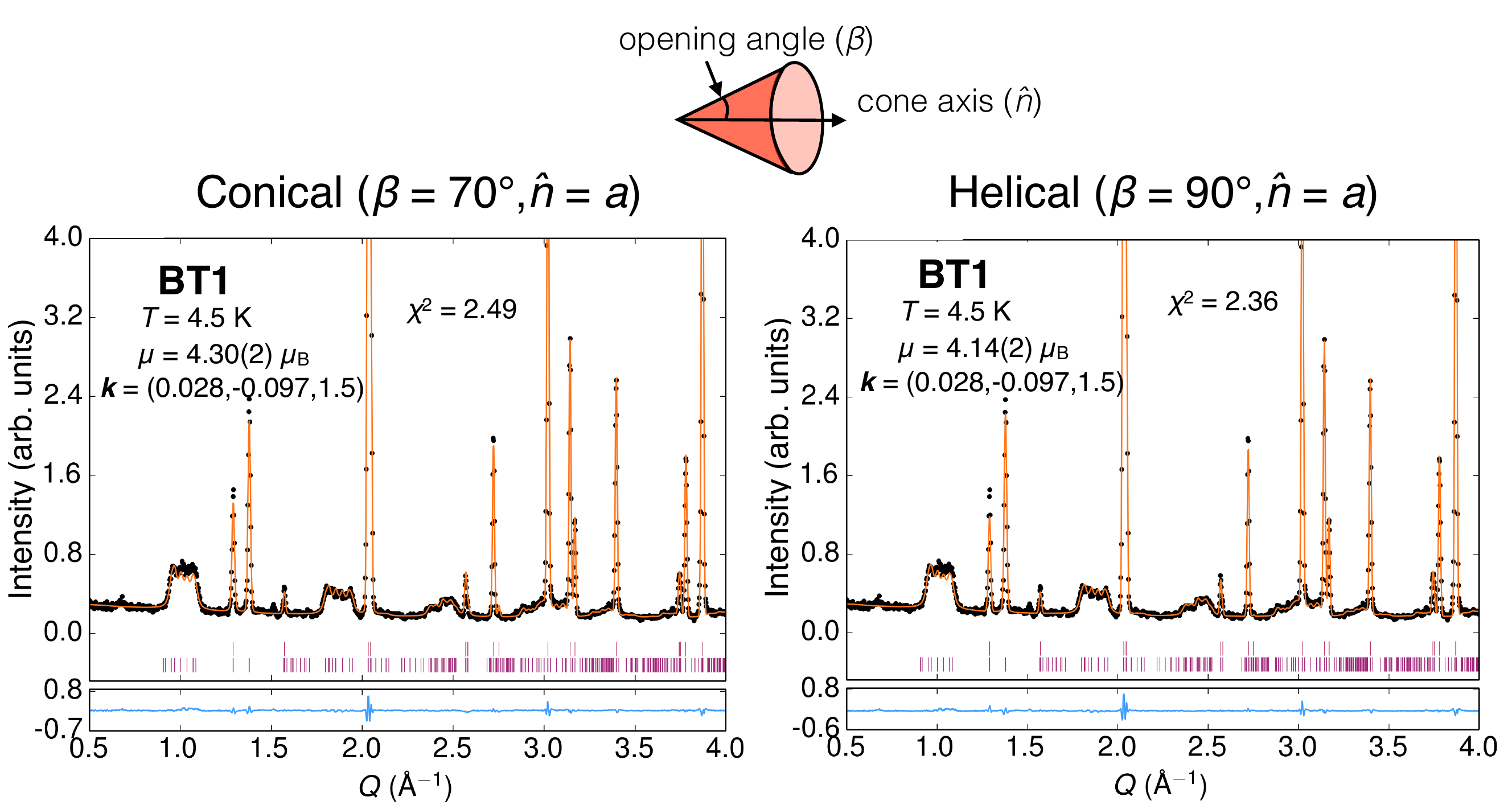}
\caption{Left: Refinement of a conical structure with opening angle 70$^{\circ}$ and cone axis along $a$, which produces a good fit to the NPD pattern.  Right: Refinement of a helical structure with opening angle 90$^{\circ}$ and cone axis along $a$ (the same refinement as shown in Fig. \ref{fig:refinement}).}
\label{fig:conical_refinement}
\end{figure}

\begin{figure}[!thb]  
\centering
\includegraphics[ width=\columnwidth]{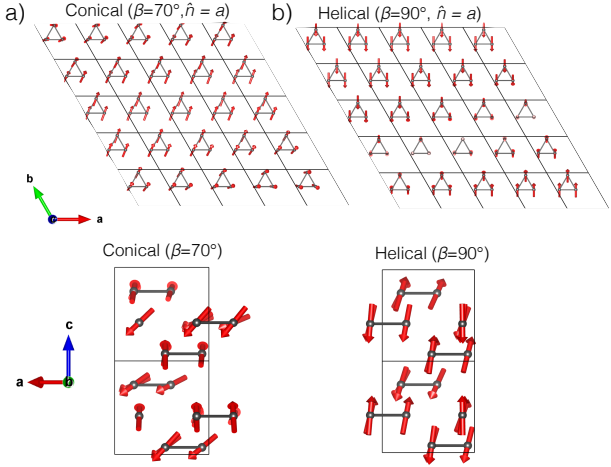}
\caption{(a) Representations of a conical structure with opening angle 70$^{\circ}$ and cone axis along $a$, which produces a good fit to the NPD pattern (top: showing only the first layer of triangles in the unit cell)  (b) Representations of the helical structure with opening angle 90$^{\circ}$ and cone axis along $a$ (the same structure as shown in Fig. \ref{fig:refinement} and discussed in the main text.)}
\label{fig:conical_struct}
\end{figure}

The NPD pattern for \fepo \ can be fit by either a helical structure or a conical structure.  These structures are related by the definition of a conical axis direction ($\hat{n}$) and an opening angle of the cone ($\beta$).  The best fits have $\hat{n}$ within the $ab$ plane, and $\beta = 70^{\circ}$ (conical) or $\beta = 90^{\circ}$ (helical).  These refinements are shown for  $\hat{n} = a$ in Fig. \ref{fig:conical_refinement} and the magnetic structures are illustrated in Fig. \ref{fig:conical_struct}.  The helical structure is the same one displayed in Fig. \ref{fig:refinement} and discussed in the text.   Note that neither structure produces a net moment.  Furthermore, in the conical structure, the moments connected by $J_2$ form an angle of $\sim$ 140$^{\circ}$ with one another, while those connected by $J_2$ within the helical structure form an angle of $\sim$ 175$^{\circ}$.  Therefore an isotropic AFM $J_2$ interaction favors helical structure over the conical one.

\section{Full $\chi$ vs. $T$ for \fepo}
\label{sec:fullchi}

\begin{figure}[!tb]  
\centering
\includegraphics[ width=\columnwidth]{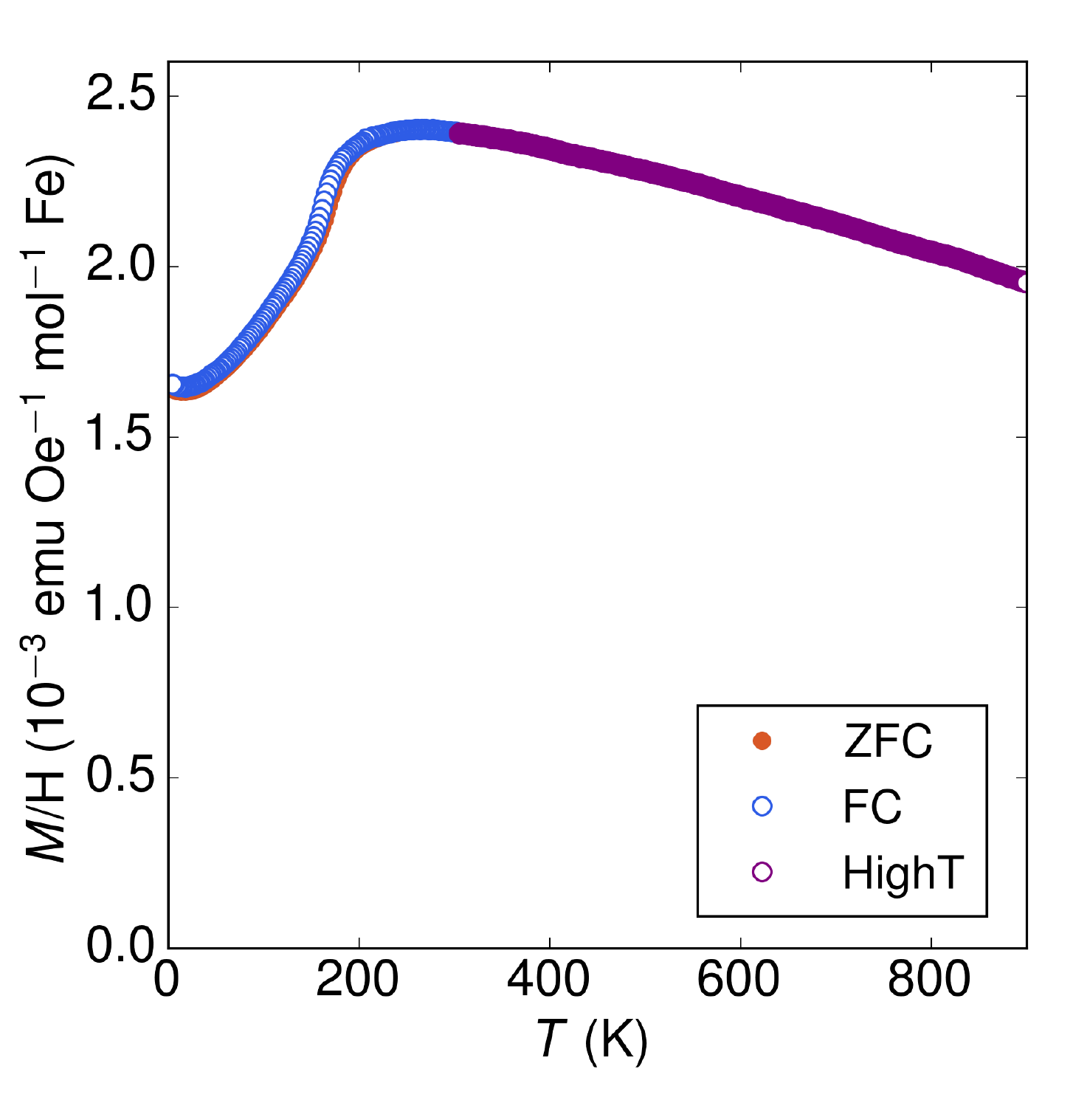}
\caption{Susceptibility vs. temperature in a 10000 Oe field for \fepo \ from 1.8 K to 900 K.  Measurements conducted under different settings are shown in different colors.  Purple: High temperature susceptibility (300 K - 900 K), shifted by +0.00011 emu Oe$^{-1}$molFe$^{-1}$. Orange: zero-field cooled low temperature susceptibility.  Blue: field cooled susceptibility.}
\label{fig:fullchi}
\end{figure}

The magnetic susceptibility of \fepo, measured under an applied field of 1 T, is shown in Figure \ref{fig:fullchi}.  The high temperature (300 K to 900 K) and low temperature (1.8 K to 300 K) data are shown in different colors, since the measurements were collected in a different sample holder.  The high temperature data was collected on a sample that was fixed to a heater using Zircar cement.  The low temperature measurement was conducted on a sample encapsulated in a plastic holder.  To achieve quantitative agreement in the overlapping region, the high temperature $M$/H was shifted by +0.00011 emu Oe$^{-1}$mol$^{-1}$ (5.5\%).  The lower absolute susceptibility obtained from the high temperature data (as compared to the lower temperature data) may be attributable to a diamagnetic contribution from the Zircar cement.

\section{Isothermal magnetization in \fepo}
\label{sec:isothermal}

\begin{figure}[!tb]  
\centering
\includegraphics[ width=\columnwidth]{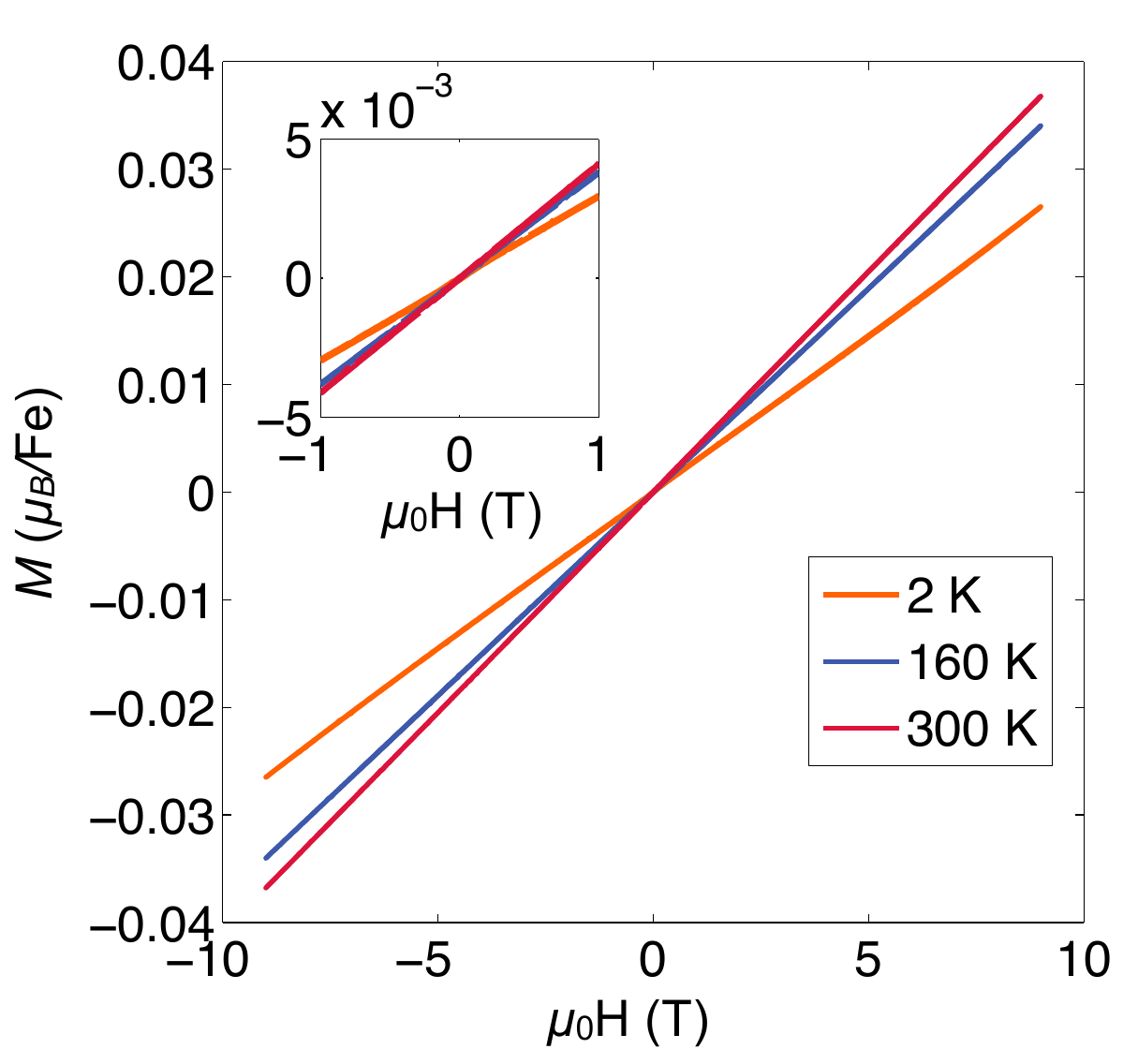}
\caption{Isothermal magnetization at $T$ = 2 K, 160 K, and 300 K.  A full field-sweep (0T $\rightarrow$ 9T $\rightarrow$ -9T $\rightarrow$ 9T) at each temperature is shown.}
\label{fig:MvsH}
\end{figure}

The magnetization of \fepo \ was measured as a function of magnetic field at several temperatures from 300 K down to  2 K.  At each temperature the field was increased from zero to 9 T, decreased to -9 T, and then increased back to 9 T.  All repeated parts of the $M$ vs. $\mu_0$H curve follow the same line, indicating no observable hysteresis at any measured temperature.    The $M$ vs. $\mu_0$H curves for three temperatures are shown in Fig. \ref{fig:MvsH}.  It is noteworthy that the curves are linear up to 9 T at all temperatures, and only a small fraction of the full moment is observed (4.2 $\mu_B$ expected at saturation, and $\sim$ 0.04 $\mu_B$ observed at 300 K and 9 T).  These observations are consistent with very strong antiferromagnetic correlations, as inferred based on the absence of Curie-Weiss behavior up to 900 K, as well as the diffuse magnetic scattering seen in neutron diffraction at 300 K (see main text).

%\bibliography{Fe3PO7.bib}

\end{document}